\documentclass[a4paper, twocolumn, accepted=2024-03-16]{quantumarticle}
\pdfoutput=1
\usepackage[utf8]{inputenc}
\usepackage[english]{babel}
\usepackage[T1]{fontenc}
\usepackage{adjustbox}
\usepackage{braket}
\usepackage{amssymb}
\usepackage{amsmath}
\usepackage{bm}
\usepackage{mathdots}
\usepackage{graphicx}
\usepackage{mathtools,amssymb,amsmath, amsthm,stmaryrd,thmtools,braket}
\usepackage[numbers,sort&compress]{natbib}
\usepackage{xr-hyper}
\usepackage[colorlinks=true, urlcolor=blue,citecolor=blue,anchorcolor=blue]{hyperref}

\usepackage{tikz}
\usetikzlibrary{quantikz}

\newcommand{\Eq}[1]{Eq.~(\ref{#1})}

\newcommand{\Fig}[1]{Fig.~\ref{#1}}

\newcommand{\ua}{\mathord{\uparrow}}
\newcommand{\da}{\mathord{\downarrow}}
\newcommand{\Ztwo}{{\mathbb{Z}_2}} 

\begin{document}
\title{Randomized measurement protocols for lattice gauge theories}
\author{Jacob Bringewatt}
\email{jbringew@umd.edu}
\affiliation{Joint Center for Quantum Information and Computer Science, NIST/University of Maryland, College Park, Maryland 20742, USA}
\affiliation{Joint Quantum Institute, NIST/University of Maryland, College Park, Maryland 20742, USA}
\author{Jonathan Kunjummen}
\email{jkunjumm@umd.edu}
\affiliation{Joint Center for Quantum Information and Computer Science, NIST/University of Maryland, College Park, Maryland 20742, USA}
\affiliation{Joint Quantum Institute, NIST/University of Maryland, College Park, Maryland 20742, USA}
\author{Niklas Mueller}
\email{niklasmu@uw.edu}
\affiliation{InQubator for Quantum Simulation (IQuS), Department of Physics, University of Washington, Seattle, WA 98195, USA.}
\begin{abstract}
\noindent Randomized measurement protocols, including classical shadows, entanglement tomography, and randomized benchmarking are powerful techniques to estimate observables, perform state tomography, or extract the entanglement properties of quantum states. While unraveling the intricate structure of quantum states is generally difficult and resource-intensive, quantum systems in nature are often tightly constrained by symmetries. This can be leveraged by the 
 symmetry-conscious randomized measurement schemes we propose, yielding clear advantages over symmetry-blind randomization such as reducing measurement costs,
 enabling symmetry-based error mitigation in experiments,  allowing differentiated measurement of (lattice) gauge theory entanglement structure, and, potentially, the verification of topologically ordered states in existing and near-term experiments. Crucially, unlike symmetry-blind randomized measurement protocols, these latter tasks can be performed without relearning symmetries via full reconstruction of the density matrix.
\end{abstract}
\maketitle

\section{Introduction}
Measurement in quantum mechanics reveals very limited information regarding the structure of the underlying quantum state. This has major  practical implications, e.g., for variational near-term quantum-classical algorithms~\cite{peruzzo2014variational,kandala2017hardware,kokail2019self,tilly2022variational}, the verification of quantum devices~\cite{eisert2020quantum}, or when detecting entanglement~\cite{friis2019entanglement} in quantum simulation experiments.  
Randomized measurement protocols, such as randomized benchmarking~\cite{knill2008randomized},
classical shadows~\cite{ paini2019approximate,huang2020predicting, huang2021efficient,hu2021classical, zhao2021fermionic,kunjummen2021shadow, levy2021classical,helsen2021estimating, huang2022quantum,hao2022classical, huang2022learning}, and 
entanglement tomography~\cite{pichler2016measurement, dalmonte2018quantum,elben2018renyi,vermersch2018unitary, elben2019statistical, brydges2019probing, elben2020mixed, zhou2020single,neven2021symmetry, kokail2021entanglement, rath2021importance, kokail2021quantum,elben2023randomized,zache2022entanglement} are valuable techniques for addressing this  problem. They allow one to estimate many observables from a few measurements~\cite{huang2020predicting,huang2021efficient} or extract non-linear quantities, such as purities $\sim \text{Tr}(\rho^k)$, $k\ge 2$ and entanglement entropies~\cite{van2012measuring, elben2018renyi,brydges2019probing,vermersch2018unitary, elben2019statistical}, potentially
without the massive overhead of traditional state tomography~\cite{flammia2012quantum, haah2016sample, o2016efficient}, see e.g.~\cite{elben2023randomized} for a recent overview. Many  techniques are feasible on noisy, near-term quantum devices~\cite{senrui2021robust, koh2022classicalshadows, tran2022measuring}.  

A key application for quantum computing  and randomized measurement protocols is simulating quantum many-body systems, with digital or analog devices based on atomic, molecular and optical (AMO), and solid-state systems~\cite{blatt2012quantum,bloch2012quantum, gross2017quantum, schafer2020tools, bassman2021simulating,monroe2021programmable, daley2022practical}. Quantum simulation promises to address long standing questions in condensed matter, high energy physics and nuclear physics. Examples include simulating non-equilibrium evolution and thermalization~\cite{deutsch1991quantum,srednicki1994chaos,rigol2008thermalization,deutsch2013microscopic,khemani2014eigenstate,eisert2015quantum,kaufman2016quantum,berges2020thermalization,zhou2022thermalization,mueller2022thermalization}, thermal systems~\cite{lu2020structure,brenes2020multipartite} and quantum phases~\cite{osterloh2002scaling,vidal2003entanglement,verstraete2004entanglement,costantini2007multipartite,li2008entanglement}.  One important frontier is the study of lattice gauge theories (LGTs)~\cite{byrnes2006simulating,banerjee2012atomic,zohar2013simulating, zohar2013quantum,zohar2013cold,tagliacozzo2013simulation,zohar2015quantum,martinez2016real, yang2016analog,zache2018quantum, klco2018quantum,lu2019simulations, barbiero2019coupling,lamm2019general,davoudi2020towards,surace2020lattice,luo2020framework,banuls2020simulating,mil2020scalable,Paulson:2020zjd,Chakraborty:2020uhf, shaw2020quantum,magnifico2020real, klco20202,Klco:2021lap,homeier2021z, Pederiva:2021tcd,zhou2022thermalization,Rajput:2021khs,nguyen2022digital, de2021quantum,rahman20212,haase2021resource,kan2021lattice,davoudi2021search,Ciavarella:2021nmj,Alam:2021uuq,Ciavarella:2021lel,cohen2021quantum, gonzalez2022hardware,halimeh2022gauge, andrade2022engineering, atas2022real,farrell2022preparations,Murairi:2022zdg,clemente2022strategies,davoudi2022quantum,mueller2022quantum,davoudi2022toward,Kane:2022ejm,mildenberger2022probing,Gustafson:2023swx,Zache2023fermion}  with intricate entanglement structures~\cite{buividovich2008entanglement, casini2014remarks, aoki2015definition, ghosh2015entanglement, van2016entanglement, lin2020comments, marco2021entanglement, mueller2022thermalization, panizza2022entanglement} and, potentially, emergent topological phases that have applications in topological quantum computation~\cite{c1982stormer,wen1990topological,kitaev2003fault,kitaev2006anyons,sarma2006topological,nayak2008non,sarma2015majorana,lahtinen2017short}.

\begin{figure*}[t]
  \centering
\includegraphics[width=0.97\textwidth]{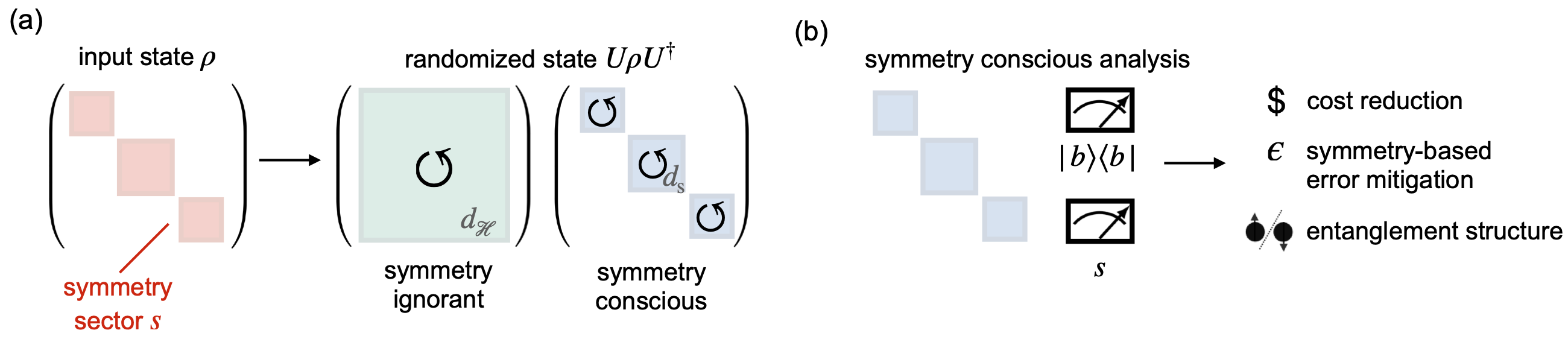}
  \caption{(a) Illustration of our proposed symmetry-conscious randomized measurement scheme, which preserves the symmetry structure of states, compared to symmetry-blind approaches. (b) Applications and benefits include cost reduction when measuring entanglement and finding classical representations of quantum states, allowing a rudimentary symmetry-based error mitigation scheme, and enabling symmetry-resolved  measurement of entanglement structure.\label{fig:overview1}}
\end{figure*}
Randomized measurement protocols are based on changing the basis via unitaries $U$ drawn from an appropriate ensemble $\mathcal{E}$, i.e. $\rho \rightarrow U \rho U^\dagger$, followed by measurement in this basis, and classical post-processing or 
quantum variational techniques to determine quantities of interest.\footnote{Schemes using collective measurement on many copies of  $\rho$ (e.g. shadow tomography~\cite{aaronson2018shadow, aaronson2019gentle}) do not fit into this class of randomized measurement protocols.}
The effectiveness and cost of a scheme depends on the choice of $\mathcal{E}$ and the particular quantities one wants to compute.
For instance, estimating expectation values, $\text{Tr}[{O_m \rho}]$, of $M$ $q$-local operators $\{O_m\}_{m=1}^M$ can be done with qubit-local random rotations, $\mathcal{O}(3^q\log(M))$ measurements, and efficient classical postprocessing~\cite{huang2020predicting}.
For non-linear quantities, i.e. those that depend on $\rho^k$ ($k\ge 2$), 
one approach is to choose an ensemble $\mathcal{E}$ that forms an (approximate) unitary $k$-design~\cite{elben2018renyi, vermersch2018unitary}.

A fundamental problem is that standard randomized measurement protocols do not account for the symmetry structure of states, e.g., by randomizing unnecessarily over unphysical bases. In particular, a (classically known) symmetry $\hat{S}$ of $\rho=\bigoplus_s \rho_s$  ($[\hat{S},\rho]=0$), where $s$ labels $\hat{S}$-eigensectors, is lost after randomization. 
Therefore, it is advantageous to perform symmetry-conscious randomized measurement by using randomizing unitaries such that $[U,\hat{S}]=0$ for all $U\in\mathcal{E}$. 
This information is compactly summarized in Fig.~\ref{fig:overview1}, 
comparing symmetry-ignorant versus the symmetry-conscious random circuits $U$  proposed in this manuscript. Our goal is to systematically study the construction of symmetry-conscious randomizing circuits and their use cases for exploring physical phenomena.  The primary results and findings of our work are as follows
\begin{enumerate}
    \item[(a)] In Section~\ref{sec:globalsymmetry}, we discuss a comprehensive approach to symmetry-conscious randomization for qubit-based models with inherent symmetries. We discuss one application, symmetry-conscious unitary $k$-designs, as an approach to global randomization. This offers several advantages, including a significant reduction in measurement complexity and allowing for symmetry-based error mitigation~\cite{nguyen2022digital}.
    \item[(b)] In Section~\ref{s:gaugetheories}, we focus on a concrete LGT example to demonstrate the application of symmetry-conscious unitary $k$-design based randomization. Our main finding is that symmetry-conscious designs enable the simultaneous measurement of both the distillable and symmetry components of entanglement, a capability that symmetry-ignorant designs lack.
    \item[(c)] In Section \ref{sec:2dz2}, we present our key result,  a protocol designed to detect topological order (TO) experimentally by assessing the gap of the entanglement spectrum (ES).  The ES is a presentation of a state in terms of an entanglement Hamiltonian in accordance with Li and Haldane's conjecture for TO states~\cite{li2008entanglement}. This has attracted significant attention in recent years, but has so far remained within the realm of theoretical exploration. 
    To address this, we leverage symmetry-conscious random measurements in combination with entanglement Hamiltonian tomography techniques. This approach represents a promising step towards experimental identification
    of topological phases which have  already started to materialize in experiments~\cite{satzinger2021realizing,semeghini2021probing}.

    The realization of an approximate unitary $k$-design is not necessary for our protocol, but it is a convenient approach to illustrate the scheme given that Haar-random measurement channels can be easily inverted. Any tomographically complete symmetry-respecting scheme could serve the same purpose.
\end{enumerate}
 Finally, in Section~\ref{s:conclusion}, we summarize our results and further discuss applications and extensions. The manuscript is supplemented by several appendices where we discuss various details of the employed numerical techniques.

\section{Symmetry-conscious randomization}\label{sec:globalsymmetry}
We begin with a general description of symmetry-conscious unitary circuit construction---i.e., circuits that preserve the symmetry structure of an input state $\rho = \bigoplus_s \rho_s$, $U= \bigoplus_s U_s$. 
Our approach, partly inspired by Refs.~\cite{elben2018renyi,vermersch2018unitary}, uses the fact that symmetry-conscious randomization can be performed using local generators that are present in the Hamiltonian of a given system. The first step is to identify local $q$-qubit generators ($1\le q \le m$) for symmetry-conscious Haar random unitaries from amongst the $m$-body terms of the relevant physical Hamiltonian $H$. 
For pragmatic reasons, such terms are good candidates for experimentally-implementable symmetry-conscious randomization: if an experimentalist can engineer the local Hamiltonian terms for the purposes of (Trotterized) time evolution, with adequate control of coupling strengths, they can also perform randomization generated by such terms. The next step
is to ensure that such terms can generate (Haar-)random unitaries over an $m$ qubit Hilbert space within every symmetry sector. The goal is to use this $m$-local symmetry-conscious randomization as a building block for randomization over the full Hilbert space. General $U(2)$ rotations are in either a `ZXZ' or `ZYZ' generator decomposition with angles $\alpha$, $\beta$ and $\gamma$ (and a global phase). In an $m$-qubit circuit one must identify the corresponding operators that act on $q$ qubits and which `embed' 1-qubit rotations in every symmetry sector. 

Different possible families of circuits can be generated by different choices of arranging these blocks within a larger circuit; the $m$ qubits interacting via a $m$-local Hamiltonian term, need not match the same (typically, geometrically local) $m$ qubits in the corresponding Hamiltonian.  While we focus primarily on generating families of unitary circuits that form symmetry-conscious unitary $k$-designs, one could also consider shallower circuits, e.g., for measuring local observables. 

To leverage global Haar randomness, we determine if the selected local, symmetry-respecting terms are sufficient to generate global, symmetry-respecting Haar random unitaries within each symmetry sector. This is not generically the case~\cite{hao2022classical}.
Verifying that sufficient randomization is, indeed, possible with the selected set of terms must be done on a case-by-case basis and accounts for the primary challenge in extending this approach to new systems. 
Here, we will demonstrate that such local terms are sufficient to generate approximate unitary $k$-designs for a few different examples of interest (with an emphasis on LGTs).

\subsection{A symmetry-conscious unitary $k$-design example: particle number symmetry}
We illustrate our approach first for spin systems with particle number symmetry, $\hat{S}_{N}\equiv\sum_j (\sigma_j^z+1)/2$, where $j$ labels a lattice site and $\sigma^a_j$ ($a=x,y,z$) are Pauli matrices acting on the site.  We emphasize that symmetry-conscious randomization schemes for particle number symmetry have been considered before~\cite{hao2022classical,zhao2021fermionic,wan2022matchgate}, but it serves as a simplified setting in which to develop the particular approach we will take to symmetry-conscious randomized measurement protocols. The lessons we learn here can then be extended to LGTs.

Symmetry-conscious randomization is achieved with components consisting of ($m$=2)-qubit gates of the form
\begin{align}\label{eq:2qubitbasicgate}
    u\equiv \begin{pmatrix}
     e^{i\theta} & &\\
     & \begin{bmatrix}
  u_1
  \end{bmatrix} & \\
  & & e^{i \phi}
    \end{pmatrix}\,,
\end{align}
where the rows and columns label (from top down) the $\da\da$ (0-particle), $\ua\da$, $\da\ua$ (1-particle) and $\ua\ua$ (2-particle) sectors; $[u_1]$ indicates the $2\times 2$ ($ZYZ$-decomposition) matrix structure,
\begin{align}\label{eq:1qubbitrotation}
[u_1]\equiv   
\begin{pmatrix}
 e^{i(\alpha+\gamma)}\cos(\beta) & e^{-i(\alpha-\gamma)}\sin(\beta) \\
- e^{i(\alpha-\gamma)}\sin(\beta) & e^{-i(\alpha+\gamma)}\cos(\beta) 
\end{pmatrix}\,.
\end{align}
If $\alpha$, $\beta$, and $\gamma$ are selected such that $[u_1]$ is drawn from a circular unitary ensemble (CUE)~\cite{brydges2019probing},  $\theta$ and $\phi$ are drawn evenly from $[0,2\pi)$, then \Eq{eq:2qubitbasicgate} is a  block-structured unitary acting Haar-randomly on the blocks of equal particle number.\footnote{An overall phase of the $2\times2$ CUE matrix was re-expressed in the 0-, and 2-particle sectors by partially absorbing it into a global phase.}  It has the following circuit realization:
\begin{equation}\label{eq:circuit}
\begin{adjustbox}{width=0.48\textwidth}
\begin{quantikz}[transparent, row sep=0.12cm]
\lstick{$j_1$}\qw & \gate{e^{\;i\frac{\alpha}{2} \sigma^z} \;} \qw & \gate[2]{\mathcal{U}(\beta)}\qw & \gate{e^{\;i\frac{\gamma}{2} \sigma^z}\;}  &\ctrl{1} \qw&\gate{\sigma^x} &\ctrl{1} \qw&\gate{\sigma^x} \\
\lstick{$j_2$} \qw & \gate{e^{-i\frac{\alpha}{2} \sigma^z}} \qw & \qw & \gate{e^{-i\frac{\gamma}{2} \sigma^z}} &\gate{P(\phi)}&\gate{\sigma^x}& \gate{P(\theta)}&\gate{\sigma^x} 
\end{quantikz}
\end{adjustbox}
\end{equation}
 where $\mathcal{U}_{j_1,j_2}(\beta)\equiv \exp\{i\frac{\beta}{2}( \sigma^y_{j_1} \sigma^x_{j_2} - \sigma^x_{j_1} \sigma^y_{j_2})\}$ and $P(x)= \text{diag}(1,\exp\{ix\})$ is the phase gate. 
 
In line with our general approach, the intuition behind \Eq{eq:circuit} comes from inspecting a particle number conserving Hamiltonian: $\mathcal{U}_{j_1,j_2}(\beta)$ is essentially a hopping term between $j_1$ and $j_2$ with amplitude $\beta/2$, the $z$-rotations before and after are density-density interactions with amplitudes $\alpha/2$ and $\gamma/2$, respectively. Together, these generate $SU(2)$ in the 1-particle block. The relative phases $\theta$ and $\phi$ are because we wish to embed $U(2)$, not $SU(2)$, and because the 0- and 2- particle blocks should be $U(1)$-randomized. (A global phase is irrelevant.)
\begin{figure}
  \centering
  \includegraphics[scale=0.33]{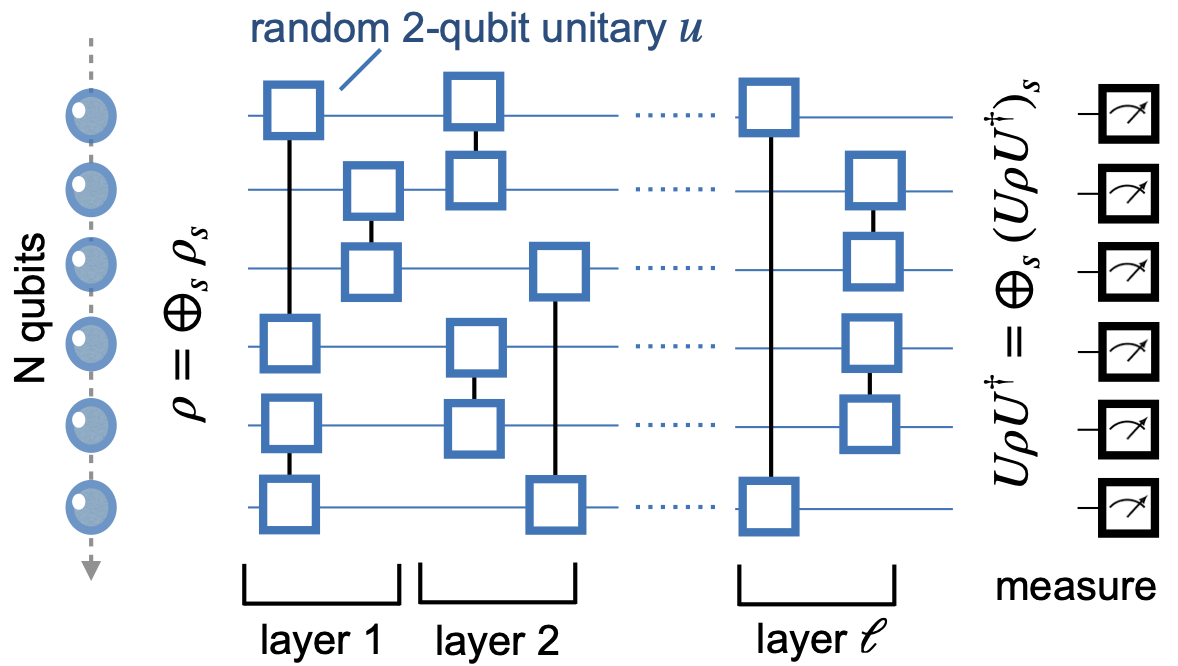}
 \caption{Random circuit scheme for systems with particle number symmetry. Blue squares connected by black lines represent the unitary $u$ in \Eq{eq:circuit}.
  \label{fig:particle_number_circuit}}
\end{figure}

As a first application, we now focus on global randomization, arranging these components into approximate unitary $k$-designs. To do so, we construct an ensemble $\mathcal{E}$ of $N_{\mathcal{E}}$ $n$-qubit symmetry-conscious random circuits, as shown in \Fig{fig:particle_number_circuit}, consisting of $\ell$ layers where $n(n-1)/2$ pairs of qubits ($(n-1)(n-2)/2$ if $n$ is odd) are randomly connected by \Eq{eq:2qubitbasicgate}. 
For sufficiently large $\ell$ and $N_{\mathcal{E}}$,
this realizes an approximate unitary $k$-design in every symmetry block $s$.

We verify that such circuits yield an approximate $k$-design numerically for $k=2$ by considering the following moments, separately for every $s$,
\begin{align}\label{eq:2designquant}
    &(\mathcal{B}^s)^{ i'j'k'l'}_{ijkl}
    \equiv \langle U^s_{ij} U^{s*}_{i'j'}U^s_{kl} U^{s*}_{k'l'}\rangle
    \nonumber\\
    & - \frac{d_s^2}{d_s^2-1} \Big[ (\mathcal{A}^s)^{i'j'}_{ij}(\mathcal{A}^s)^{k'l'}_{kl}+ (\mathcal{A}^s)^{k'l'}_{ij}(\mathcal{A}^s)^{i'j'}_{kl}\Big]\,,
\end{align}
 where $\langle \dots 
\rangle =(1/N_{\mathcal{E}}) \sum_{U\in \mathcal{E}}\dots$ is the $\mathcal{E}$-ensemble average, $d_s$ is the dimension of sector $s$, and $(\mathcal{A}^s)_{ij}^{kl}\equiv \langle U^s_{ij}U^{s*}_{kl} \rangle$. For a 2-design it holds that
\begin{align}\label{eq:def2design}
   & (\mathcal{B}^s)^{i'j'k'l'}_{ijkl} \stackrel{\text{2\text{-des.}}}{=} 
   -\frac{ \delta_{ii'}\delta_{kk'}\delta_{jl'}\delta_{lj'} + \delta_{ik'}\delta_{ki'}\delta_{jj'}\delta_{ll'} }{d_s(d_s^2-1)}\,,
\end{align}
while for a 1-design $(\mathcal{A}^s)_{ij}^{kl}=\delta_{ij}\delta_{kl}/d_s$~\cite{collins2006integration, puchala2017symbolic}. 
 We numerically simulate an ensemble of $N_{\mathcal{E}}$ circuits and compute their deviation from a 2-design $\epsilon$ as the absolute difference between \Eq{eq:def2design} and \Eq{eq:2designquant}, averaged over indices and multiply by $d_s(d_s^2-1)$ to make $\epsilon$ dimension-independent. Up to a rescaling of the approximation ratio, this choice of quantifying the error between our random ensemble and a 2-design is equivalent to other standard choices, such as the diamond norm or the the frame potential. This is demonstrated in Appendix~\ref{app:equiv_error_metrics}.
\begin{figure}[t]
  \centering
  \includegraphics[width=\columnwidth]{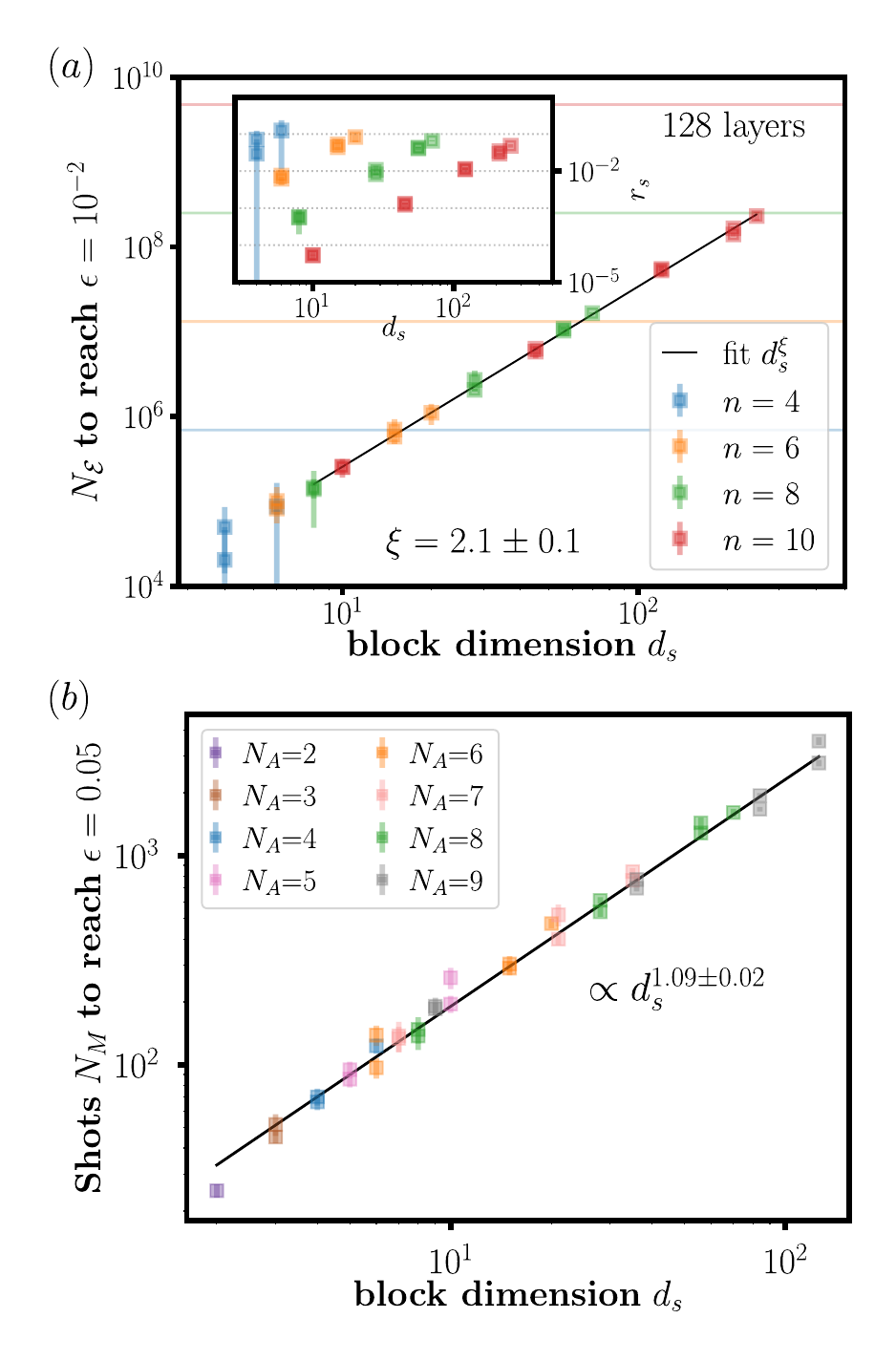}
  \caption{(a)  Particle number-respecting random circuits $N_{\mathcal{E}}$ required to approximate a 2-design with precision better than $\epsilon =10^{-2}$ in every symmetry sector ($\ell =128$) for a total number of sites $N=4,6,8,10$. Inset: Relative sampling cost reduction compared with a symmetry-ignorant scheme. (b) Measurements $N_M$  required to estimate sector-wise $2$-purities with precision better than $\epsilon=0.05$, for fixed $N_{\mathcal{E}}=1,428$ and $\ell=128$. Here, the states considered are reduced density matrices on subsystems of dimension $N_A=N/2$ (i.e. on a bipartition of the lattice.)\label{fig:scalingpn}}
\end{figure}

\Fig{fig:scalingpn}(a) displays the size $N_{\mathcal{E}}$ of a circuit ensemble (with $\ell=128$ layers) to approximate a 2-design better than $\epsilon=10^{-2}$ for all particle number blocks of $n=4,6,8,10$ qubit systems. The sampling complexity scales  with the block-dimension $d_s$ as
\begin{align}\label{eq:scaling2design}
   N_\mathcal{E}\sim d_s^{\xi} \,,\quad \xi=2.1\pm 0.2\,,
\end{align}
consistent with $\xi=2$. Error bars are found by comparing with the $\ell\rightarrow \infty$ limit (obtained by directly sampling Haar random unitaries in each symmetry block); the fit error for $\xi$ is determined by varying the fit regime, leaving out the largest few blocks, and by varying the required $\epsilon$ by one order of magnitude. 
 The inset of \Fig{fig:scalingpn}(a) shows the relative reduction $r_s$ in sampling cost compared to using a symmetry-ignorant scheme, where $r_s:=(d_s/d_{\mathcal{H}})^{\xi}$. Largest gains are found away from half filling, which can be understood by comparing $d_{\mathcal{H}}=2^n$ to $d_s = \binom{n}{s}=n!/(s!(n-s)!)$ and using Stirling's approximation to find that $r_s \sim (d_s/d_{\mathcal{H}})^\xi \approx [ 2 (\frac{s}{n})^{\frac{s}{n}}(1-\frac{s}{n})^{1-\frac{s}{n}}  ]^{-n\xi}/ (2\pi n \frac{s}{n} (1-\frac{s}{n}))^{{\xi}/{2}}$, an exponential (in $n$) cost reduction for $s \ll n/2 $ ($\gg n/2$).

A primary concern for constructing symmetry-conscious random circuits is circuit complexity---in particular, the required number of layers $\ell$ needed to reach a given precision as a function of the system size. To investigate this, we conducted numerical simulations on systems involving up to $n$=10 qubits. In Appendix \ref{app:numerical_details}, we provide a detailed account of the number of layers necessary to represent a $k$-design (specifically a 2-design), comparing our findings with exact Haar random sampling results. For the system sizes we can consider numerically this convergence is very rapid. For instance, for $n=10$ qubits with $N_\mathcal{E}=8192$ random unitaries the difference between sampling from our circuits versus sampling directly from the CUE is already very small within $\ell\approx 15$ layers for all symmetry sectors. The exact value of this error floor is set by $N_\mathcal{E}$, falling off with the expected $\sim 1/\sqrt{N_\mathcal{E}}$ scaling, see Appendix~\ref{app:numerical_details}.
As our numerical methods are restricted to relatively small systems and considering that $k$-designs serve as one application rather than the core focus of our work, we refrain from asserting an exact analytical form of the error scaling that holds true for large $n$. Nonetheless, it seems reasonable to conclude from our data that the behavior scales as a low-degree polynomial, and that symmetry-conscious circuits require comparable depths when compared to symmetry-agnostic circuits. For convenience, we continue to work in the large layer limit, in practice $\ell =128$, thus massively overdoing the actual number of layers needed.  We will return to analytic estimates in future work.

\subsection{Estimating $k$-purities from $k$-designs}
Next, we explore measuring $k$-purities using a prototypical model with particle number symmetry; we consider the following spin Hamiltonian in $(1+1)$d,
\begin{align}\label{eq:HamiltonianModel1}
    H = \frac{1}{2a}\sum_{j=0}^{N-1}( \sigma^+_j\sigma^-_{\scriptsize j+1} + {\rm h.c.}) + m\sum_{j=0}^{N-1}(-1)^j \sigma^+_j \sigma^-_j\,,
\end{align}
on a lattice with $N$ sites labelled by $j\in[0,N-1]$ and with periodic boundary conditions (PBCs);
 $\sigma_j^\pm \equiv\frac{1}{2} (\sigma_j^x\pm i\sigma_j^y)$, $a$ is the lattice spacing and $m$ a mass term. \Eq{eq:HamiltonianModel1} conserves particle number, $[H,\sum_j \frac{1}{2} (\sigma^z_j+1)]=0$. As inputs for the particle number symmetry-conscious scheme in \Fig{fig:particle_number_circuit}, we numerically determine ground states of $H$ via exact diagonalization~\cite{WeinbergED}.  Because these are simple, unentangled computational basis states in the limit $1/2a \to 0$, we work in the opposite limit with $m\cdot a  = 0.05$ (the ground states are half-filled).  We focus on a subsystem A with $ N_A = N/2$ sites. Symmetries of $\rho_A$ are particle number $s \equiv n_A\in[0,N_A]$ in the subsystem, noting that if particle number is fixed globally, $\rho_A$ is block-diagonal, $[\rho_A, \sum_{j=0}^{N_A-1} \frac{1}{2}(\sigma^z_j+1)]=0$, i.e.
$\rho_A =\bigoplus_s \rho_{A,s} $.

We extract $k$-purities, $\text{Tr}(\rho_{A,s}^k)$ by measuring  the probabilities $P_U(b,s)$ of bitstring $b$ (and symmetry sector $s$) with $N_M$ shots in the basis defined by the random unitary $U$; $k$-purities for $k\geq 2$ are directly related to the $k$-R\'{e}nyi entropies $S_{A,s}^{(k)}\equiv [1/(1-k)]\log\text{Tr}( \rho_{A,s}^k)$. Following the approach taken in Refs.~\cite{enk2012, elben2018renyi, vermersch2018unitary}, 
stochastic moments $\langle P_U(b,s)^k\rangle \equiv (1/N_\mathcal{E}) \sum_{U\in \mathcal{E}} P_U(b,s)^k $ are related to $k$-purities via~\cite{vermersch2018unitary},
\begin{align}\label{eq:kpurities}
    \langle P_U(b,s)^k\rangle = \frac{1}{D_k} \sum_{\{ a_i \}_k \in \mathbb{N}_0}C_{\{ a_i \}_k} \prod_{j=0}^k \text{Tr}\left[ \rho_{A,s}^j \right]^{a_j}
\end{align}
where $\{ a_i \}_k\equiv a_1,\dots,a_k\in \mathbb{N}_0$ with $\sum_{j=1}^kja_j=k$, $D_k\equiv\prod_{j=0}^{k-1}(d_s+j)$ and $C_{\{ a_i \}_k}\equiv k!/\prod_{j=1}^k( j^{a_j} a_j!  )$; $\langle P_U(b,s)^k\rangle$ and $\text{Tr}\left[ \rho_{A,s}^k \right]$ refer to the $k$-moments of the probabilities and $k$-purity per symmetry sector $s$, respectively, with $ \sum_{b\in s} P_U(s,b) = \text{Tr}[\rho_{A,s}]\equiv p_s \le 1$ and $\sum_s p_s=1$. We assume an ideal quantum machine, and the total measurement cost is $N_{M}\cdot N_{\mathcal{E}}$.

In \Fig{fig:scalingpn}(b), we show the required shot number $N_M$ to measure $S_{A,s}^{(2)}$ with precision better than $\epsilon=0.05$, for fixed $N_{\mathcal{E}}=$ 1,428 and $\ell = 128$ layers,  comparing different system sizes and symmetry sectors. The fit reveals an approximately linear dependence on $d_s$, error bars indicate standard error of the mean for seven independent trials. (In \Fig{fig:True_purities_particle_number} of Appendix \ref{app:numerical_details} we investigate the $N_{\mathcal{E}}$ dependence in the infinite shot limit.) Together, our results show that the cost of the symmetry-conscious approach, $N_{\mathcal{E}}\cdot N_{M}$, is proportional to $d_s$ instead of  $d_{A}$  (the size of the subsystem). In many cases $d_s\ll d_{A}$, yielding a significant advantage. We shall see that this advantage is exponential for lattice gauge theories.

\section{Lattice gauge theory entanglement}\label{s:gaugetheories}
\subsection{\texorpdfstring{$\Ztwo$}{Z2} LGT in (1+1)d}\label{ss:1dz2}
Lattice gauge theories (LGTs) are systems with an extensive number of local constraints in the form of Gauss laws defining a physical sub-Hilbert space. 
One of the key applications is the simultaneous measurement of distillable and symmetry components of entanglement. This capability is exclusive to symmetry-conscious designs and is not attainable through symmetry-ignorant schemes; it would otherwise require full state tomography. 
\begin{figure}[t!]
  \begin{center}
  \includegraphics[scale=0.33]{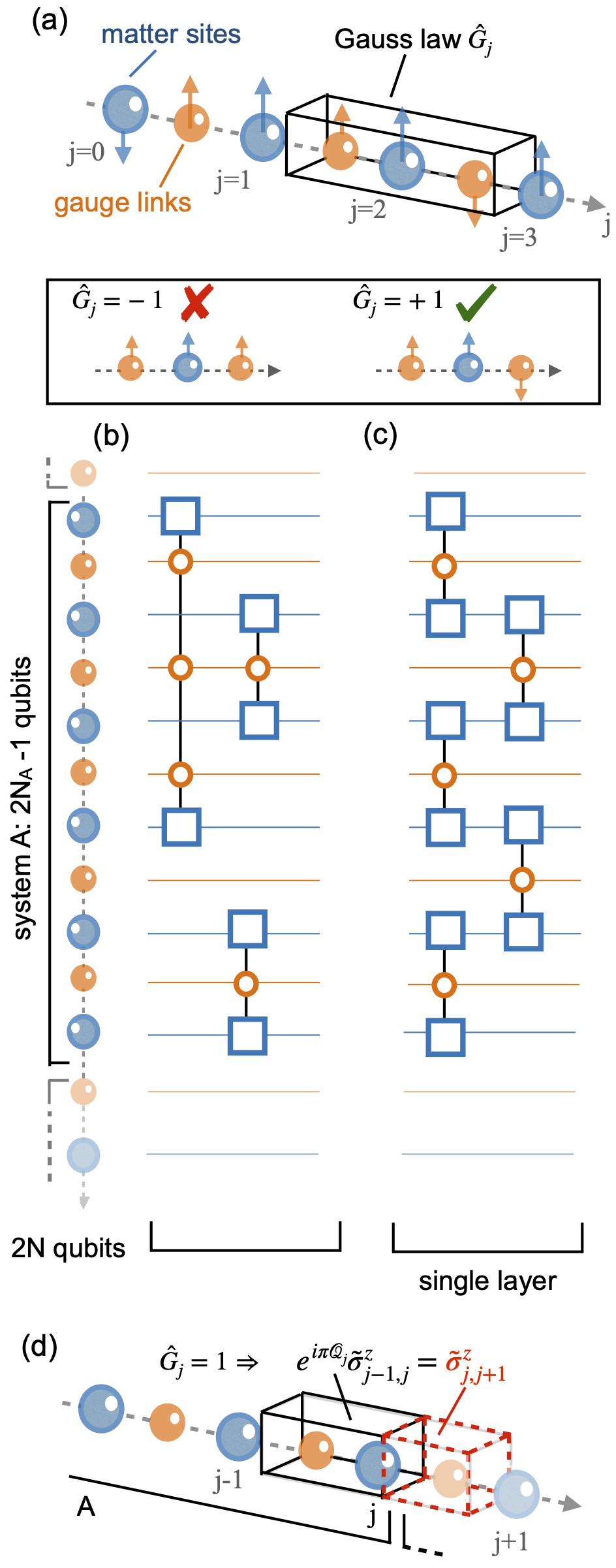}
  \caption{(a) Illustration of $\Ztwo$ LGT in $1+1$ dimensions,  depicted is an ``even'' site, $(-1)^j=1$, where spin up (down) is the presence (absence) of a $\Ztwo$ charge, $\mathcal{Q}_j=+1$ $(0)$. (b) Single layer of a random-measurement circuit with non-local gates $2+|j_1-j_2|$ qubit gates, an extension of the strategy in section \ref{sec:globalsymmetry}. (c) Near-term strategy based on  3-qubit unitaries. (d) Illustration of the symmetry structure of $\rho_A$.\label{fig:overviewLGT1}}
\end{center}
\end{figure}
We  consider $\Ztwo$ LGT coupled to staggered matter in $1+1$ dimensions, with Hamiltonian, 
\begin{align}\label{eq:Z2LGT1d}
    H = &\frac{1}{2a} \sum_{j=0}^{N-1}( \sigma^+_j \tilde{\sigma}^x_{j,j+1} \sigma^-_{j+1} + {\rm h.c.} )
    \nonumber\\
    &+ m\sum_{j=0}^{N-1}\frac{(-1)^j}{2}(1+\sigma^z_j)   + e\sum_{j=0}^{N-1} \tilde{\sigma}^z_{j,j+1},
\end{align}
Gauss laws 
\begin{equation}\label{eq:Gausslaw1d}
    \hat{G}_j \equiv e^{i\pi\mathcal{Q}_j} \tilde{\sigma}^z_{j-1,j} \tilde{\sigma}^z_{j,j+1},
\end{equation}
and periodic boundary conditions (PBC),
where $j$ labels a site and $[H,\hat{G}_j]=0$; $\sigma^{b}_j$ ($\tilde{\sigma}^b_{j,j+1}$), $b=x,z$, are Pauli operators residing on the sites (links) of the lattice and representing the matter (gauge) degrees of freedom of the theory; $m$ is a mass parameter, $e$ the $\Ztwo$ coupling, $a$ the lattice spacing, $\sigma^\pm_j \equiv (\sigma^x_j \pm i\sigma^y_j)/2$, and $\mathcal{Q}_j\equiv (\sigma^z_j+(-1)^j)/2$ is the $\Ztwo$ charge. This is compactly summarized in \Fig{fig:overviewLGT1}(a) where Gauss law eigensector with $\hat{G}_j | \psi \rangle = +1| \psi\rangle$ are physical.  
 \begin{figure*}[t]
  \centering
  \includegraphics[scale=0.47]{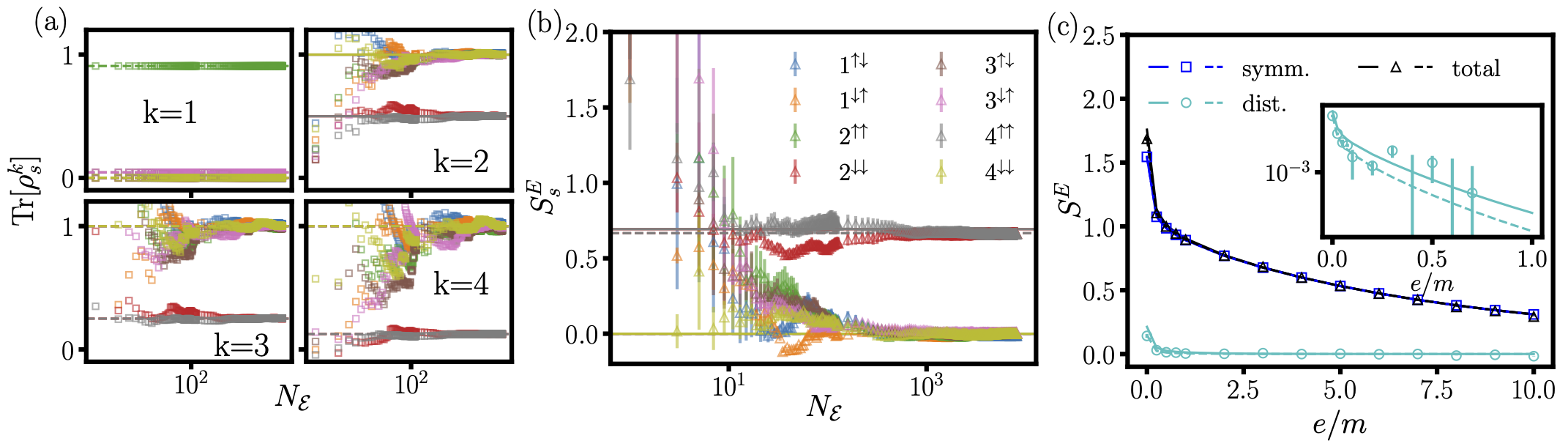}
  \caption{(a) Sector-wise $k$-purities of the $\mathbb{Z}^{1+1}_2$ ground state measured based on \Eq{eq:kpurities} as a function of $N_\mathcal{E}$, for $N_A=N/2=5$, $m\cdot a=0.1$, $e/m=8$ and $\ell=32$. (b) Von Neumann Entropy from $k$-purities/R\'{e}nyi entropies using a $4^{\rm th}$ order finite-difference approximation of \Eq{eq:vonNuenanDef}. (c) Symmetry- (symm.), distillable  (dist.) and total entanglement entropies as a function of $e/m$ for  $N_\mathcal{E}$=2000, $m\cdot a=0.1$ and $\ell=32$. Symbols represent random measurement results, dotted lines represent approximating $S_E$ by (exact) R\'{e}nyi-entropies up to $k=4$, solid lines are exact results. \label{fig:LGT1dInnsbruckScheme}}
\end{figure*}

\Fig{fig:overviewLGT1}(b) and (c) depict random measurement strategies for this model, we investigate a $N_A$ site subsystem with boundary conditions ending in a matter site. The first strategy, (b), is an extension of the circuit in Section \ref{sec:globalsymmetry}, \Eq{eq:circuit}, made gauge invariant by introducing Wilson lines connecting sites $j_1$ and $j_2$,   
\begin{align}
    W_{j_1,j_2} \equiv \prod_{j=j_1}^{j_2-1}\tilde{\sigma}^x_{j,j+1}
\end{align}
 and with circuit representation,
 \begin{equation}\label{eq:circuit_gauge}
\begin{adjustbox}{width=0.49\textwidth}
\begin{quantikz}[transparent, row sep=0.22cm]
\lstick{$j_1$}\qw & \gate{e^{\;i\frac{\alpha}{2} \sigma^z_{j_1}} \;} \qw & \gate[3]{\mathcal{W}_{j_1,j_2}(\beta)}\qw & \gate{e^{\;i\frac{\gamma}{2} \sigma^z_{j_1}}\;}  &\ctrl{2} \qw & \gate{\sigma^x} &\ctrl{2} & \gate{\sigma^x} \\
 \qwbundle[alternate]{} & \qwbundle[alternate]{} & \qwbundle[alternate]{} &  \qwbundle[alternate]{}&\qwbundle[alternate]{}& \qwbundle[alternate]{} & \qwbundle[alternate]{}& \qwbundle[alternate]{}
\\
\lstick{$j_2$} \qw & \gate{e^{-i\frac{\alpha}{2} \sigma^z_{j_2}}} \qw & \qw & \gate{e^{-i\frac{\gamma}{2} \sigma^z_{j_2}}} &\gate{P(\phi)} & \gate{\sigma^x} & \gate{P(\theta)} &  \gate{\sigma^x}
\end{quantikz}\,,
\end{adjustbox}
\end{equation}
where $\mathcal{W}_{j_1,j_2}(\beta)\equiv \exp\{i\frac{\beta}{2} (\sigma_{j_1}^y  W_{j_1,j_2}\sigma^x_{j_2}-\sigma_{j_1}^x W_{j_1,j_2} \sigma^y_{j_2} )\}$; the angles $\alpha$, $\beta$, $\gamma$, $\phi$ and $\theta$ are randomly drawn as before.
The abbreviated middle qubit bundle in \Eq{eq:circuit_gauge} refers to qubits representing gauge links (orange in \Fig{fig:overviewLGT1}),  $j_1$ and $j_2$ start and end on matter sites (blue in \Fig{fig:overviewLGT1}). This results in a  $2+|j_1-j_2|$ qubit unitary which is not a feasible strategy near-term. Because of this we will focus on a 3-qubit unitary strategy, \Fig{fig:overviewLGT1}(c), at the cost of somewhat deeper circuits to obtain $k$-designs. We demonstrate numerically that these circuit form approximate 2-designs in Appendix~\ref{app:z21ddetails}, and focus here on measuring entanglement. 
We note that (b) and (c) generate $k$-designs in every symmetry sector, despite the fact that no explicit 2-qubit entangling operation is performed between gauge sites (orange). This is a consequence of gauge symmetry: For this simple model the gauge link degrees of freedom are not truly independent because
they could have been eliminated using Gauss' law.\footnote{A caveat is that with PBCs a gauge zero mode cannot be integrated out. A $k$-design acting on the full system, not just a subsystem,  would include a modification which depends on how the remaining bosonic mode is digitized; see e.g. \cite{becker2022classical,gu2022efficient} for bosonic random measurement schemes.}

\subsection{Measuring distillable and symmetry components of entanglement}
We now put our randomizing circuits to use by demonstrating their utility for determining the entanglement structure of ground states. In the literature, this work included, the term ``entanglement structure'' is used fairly ambiguously to denote anything characterizing entanglement beyond entanglement entropies. This includes the separation of entanglement entropies into distillable and symmetry entanglement, but, more generally, also includes the structure in terms of an entanglement Hamiltonian, eigenvalue spectrum (the so-called Schmidt spectrum), and symmetries of a reduced density operator. We will consider all of these in this section.

In particular, we consider a bipartition $\rho_A\equiv \text{Tr}_{\bar{A}}(\rho)$ of the lattice with $N_A\equiv N/2$ sites. As before, $\rho_A$ is block-diagonal in particle number $n_A$ (i.e., $[\rho_A,\sum_{j=0}^{N_A-1} (1+\sigma^z_j)/2]=0$) but additionally has  symmetries beyond those of the non-gauge spin model. They are illustrated in \Fig{fig:overviewLGT1}(d): Because of Gauss' law, on the right (left) boundary $j=j_R\equiv N_A$ ($j=j_L\equiv 0$), out-going (in-going) electric fields can be written as
\begin{align}
    \tilde{\sigma}^z_{j,j+1}=\exp\{ i\pi \mathcal{Q}_{j}\}   \tilde{\sigma}^z_{j-1,j }\,.
\end{align}
  These operators are symmetries of $\rho_A$ (i.e. $[ \tilde{\sigma}^z_{j_R,j_R+1} , \rho_A]=[ \tilde{\sigma}^z_{j_L-1,j_L} , \rho_A]=0$)  if $\rho$ is physical (i.e. Gauss' law respecting) and result in a four-block symmetry structure. Together, we label the sectors as $s\equiv n_A^{s_L,s_R}$ where $s_{L/R}=\ua/\da$ and $n_A\in [0,N_A]$ is particle number. 

 In \Fig{fig:LGT1dInnsbruckScheme}(a), we show $k$-purities, $\text{Tr}[\rho_{A,s}^k]$  reconstructed by inverting \Eq{eq:kpurities}, measured for $N=10$ ($N_A=5$), $m\cdot a=0.1$, $e/m$=8 and $\ell=32$. We plot them as a function of  $N_{\mathcal{E}}$; measurements are obtained in the infinite shot  limit ($N_M\rightarrow\infty$).  The $k=1$ result $p_s\equiv \text{Tr}[\rho_{A,s}]$  is recovered exactly by design, higher ($k\ge 2$) purities are recovered for sufficiently large $N_{\mathcal{E}}$ (data is shown up to $N_{\mathcal{E}}=2^{14}$). Not shown is the cost in $N_M$ to obtain constant error which, as in Sect.~\ref{sec:globalsymmetry},  scales with $d_s$.  In \Fig{fig:LGT1dInnsbruckScheme}(b), we  estimate the von-Neumann entropy per symmetry block,
 \begin{align}\label{eq:vonNuenanDef}
     S_s\equiv \text{Tr}_s[\bar{\rho}_{A,s}\log(\bar{\rho}_{A,s})] = - \lim_{k\rightarrow 1^+} \frac{{\rm d}}{{\rm d} k} \text{Tr}_s[\bar{\rho}_{A,s}^k],
 \end{align}
 where $\bar{\rho}_{A,s}\equiv {\rho}_{A,s}/p_s$. We make use of a $4^{\rm th}$ order finite-difference approximation of the derivative to derive \Eq{eq:vonNuenanDef} from the measured $k$-purities;
error bars are obtained from comparing a $4^{\rm th}$ and $3^{\rm rd}$ order derivative.
Finally, \Fig{fig:LGT1dInnsbruckScheme}(b) shows the decomposition of the von-Neumann entropy, $S=-\text{Tr}_A[\rho_A \log(\rho_A)]$ into a symmetry- (`classical' entanglement) and distillable component, $S^{  {\rm symm.}}+S^{\rm dist.}$, 
\begin{align}\label{eq:entcomponents}
 S^{  {\rm symm.}} \equiv -\sum_s  p_s \log(p_s)\,, \quad
 S^{\rm dist.}\equiv \sum_s p_s S_s \,,
\end{align}
 as a function of $e/m$ and for $m\cdot a=0.1$.

 Finally, we note that a difference in the number of unitaries required for approximating a 2-design versus the seemingly lower requirements for measuring 2-fidelities evident in \Fig{fig:LGT1dInnsbruckScheme}. It is important to note that the accuracy of reproducing the latter is inherently dependent on the state itself. In contrast, the analysis we perform in section \ref{sec:globalsymmetry} provides an upper bound, ensuring convergence for \textit{any} input state. 
 
\begin{figure}[t]
  \centering
  \includegraphics[scale=0.46]{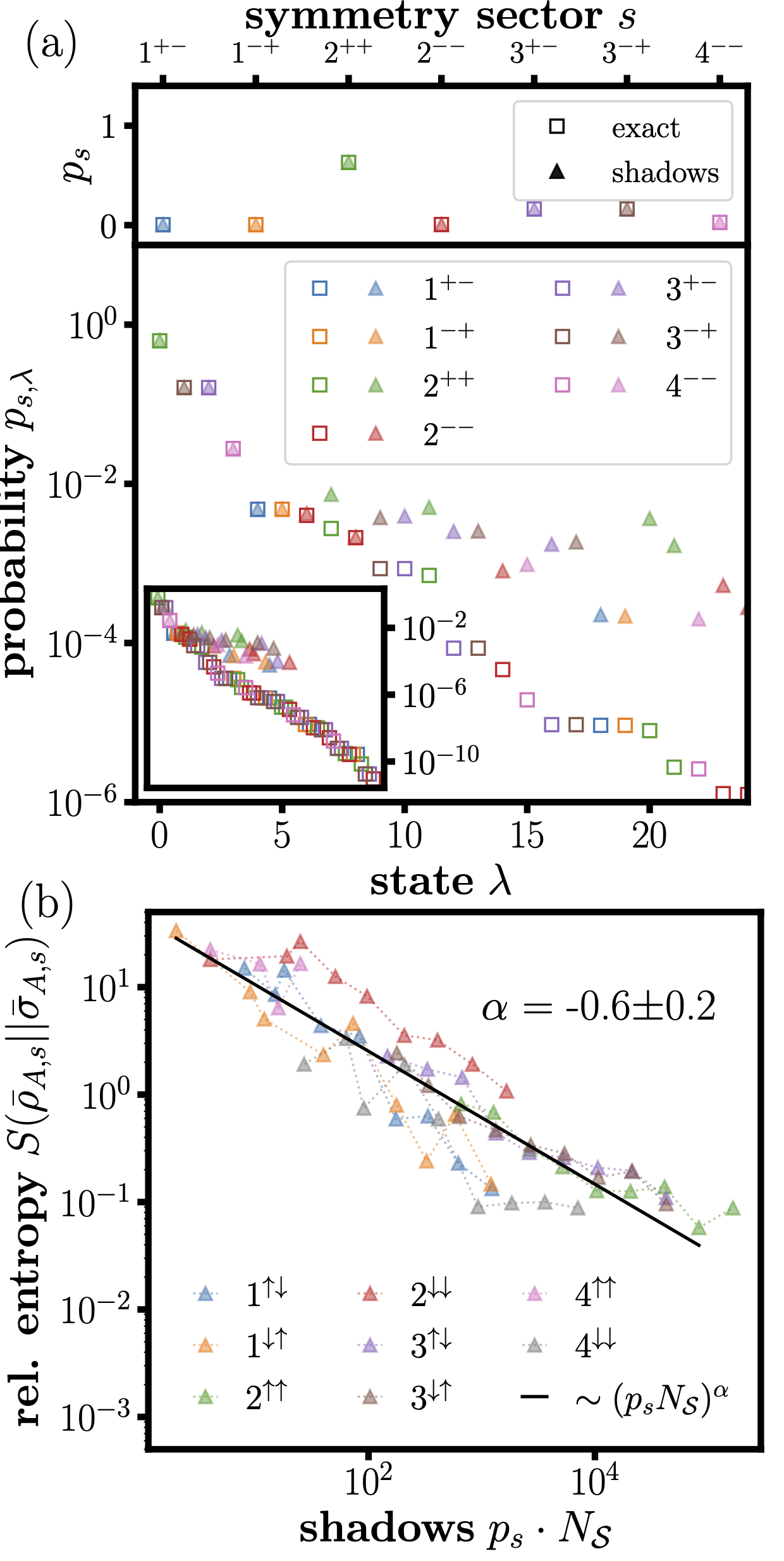}
  \caption{(a) Bottom: Shadow-reconstructed  symmetry-resolved Schmidt spectrum, $m\cdot a=0.05$, $e/m=1$, $\ell=32$ and $N_{ \mathcal{S}}=2^{17}$. Inset: Full spectrum. Top: Probability per sector $p_s$. (b) Sector-wise relative entropy $S(\bar{\rho}_{A,s}||\bar{\sigma}_{A,s})$ between  (normalized) exact $\bar{\rho}_{A,s}$ and shadow-reconstructed reduced density matrices $\bar{\sigma}_{A,s}$, as a function of shadow number $N_{\mathcal{S}}$ with $m\cdot a=0.05$, $e/m=1$ and $\ell=32$.\label{fig:shadows1d}}
\end{figure}

\subsection{Classical shadows}
Next, we  explore classical representations of $\rho_A$, starting with the classical shadow formalism of~\cite{huang2020predicting}. The basic idea is to randomize $\rho_A \rightarrow U \rho_A U^{\dagger}$ and perform a computational basis measurement yielding a bitstring $b$ from which the sector $s$ can be read off. A symmetry-conscious shadow is $U^\dagger_s | b,s \rangle \langle b,s | U_s$, where the subscript $s$ indicates that one works in block $s$ with dimension $d_s$. The ensemble of random rotations yields a CUE-random quantum channel $\mathcal{M}[\rho_{A}]=\bigoplus_s\mathcal{M}_s[\rho_{A,s}]$, and with many measurements and, consequently, many shadows one obtains classical sector-wise state representations by taking the expectation value
\begin{align}
    \bar{\sigma}_{A,s} \equiv \mathbb{E}\left[ \mathcal{M}_s^{-1} \big( U^\dagger_s | b,s \rangle \langle b,s | U_s\big)\right]\,,
\end{align}
where $\mathbb{E}[\dots]\equiv (1/N_{\mathcal{S}})\sum_{i=0}^{N_{\mathcal{S}}-1}[\dots] $ is the  $N_\mathcal{S}$ shadow average and $\mathcal{M}_s^{-1}(X)\equiv (d_s+1)X - \text{Tr}_s[X] \, \mathbb{I}_s$; $\text{Tr}_s$ and $\mathbb{I}_s$ are the sector-wise trace and identity, respectively. The bar indicates normalization, i.e. $\bar{\sigma}_{A,s}\equiv {\sigma}_{A,s}/p_s$, 
$\text{Tr}_s[\bar{\sigma}_{A,s}]=1$, $p_s\equiv \text{Tr}_s[{\sigma}_{A,s}]$ is simply the number of shadows measured in one sector relative to $N_{\mathcal{S}}$. 

 \Fig{fig:shadows1d} compactly summarizes the results of this analysis, showing the symmetry-resolved Schmidt spectrum of $\rho_A$ in (a), comparing exact results (empty squares) versus shadows (filled triangles) for $m\cdot a=0.05$, $e/m=1$, $\ell=32$ and $N_{ \mathcal{S}}=2^{17}$. The eigenvalue spectrum is well reproduced down to $P_{s,\lambda} \approx 10^{-2}-10^{-3}$; the inset shows the full spectrum and the top of (a) the accurately recovered probability for each block, $p_{s}$.
\Fig{fig:shadows1d}(b) shows the relative entropy, $S(\bar{\rho}_{A,s}|| \bar{\sigma}_{A,s})\equiv  \text{Tr}_s[\bar{\rho}_{A,s}\log(\bar{\rho}_{A,s})-\bar{\rho}_{A,s}\log(\bar{\sigma}_{A,s})]$, between exact $\bar{\rho}_{A,s}$ and  shadow-reconstructed $\bar{\sigma}_{A,s}$, as a function of shadows per sector $p_s\cdot N_{\mathcal{S}}$~\footnote{Because the average shadow density matrix at finite $N_{\mathcal{S}}$ might not be positive definite, we follow the regularization strategy of~\cite{acharya2021informationally} to project onto a positive definite matrix. The error of the scaling exponent includes an estimate of the regularization dependence, estimated by comparing with replacing $\log(\cdot) \rightarrow \frac{1}{2}\log{(\cdot^2)}$.}. The fit (black curve) indicates approximate power law $\sim (p_s N_{\mathcal{E}})^{-\frac{1}{2}}$ convergence. We note that automatically fewer shadows are  sampled in less important sectors.

While $\mathbb{Z}_2^{1+1}$ LGT serves as a useful case study, we next consider $2+1$ dimensions where our ability to extract LGT entanglement structure provides a window to studying topologically ordered systems. We add that symmetry-conscious shadows  based on deep scrambling $k$-designs are not ideal for estimating local observables, a single layer scheme is better suited for this task.
 
\section{Experimental verification of topological phases}\label{sec:2dz2}
 \begin{figure*}[t]
  \centering
  \includegraphics[scale=0.29]{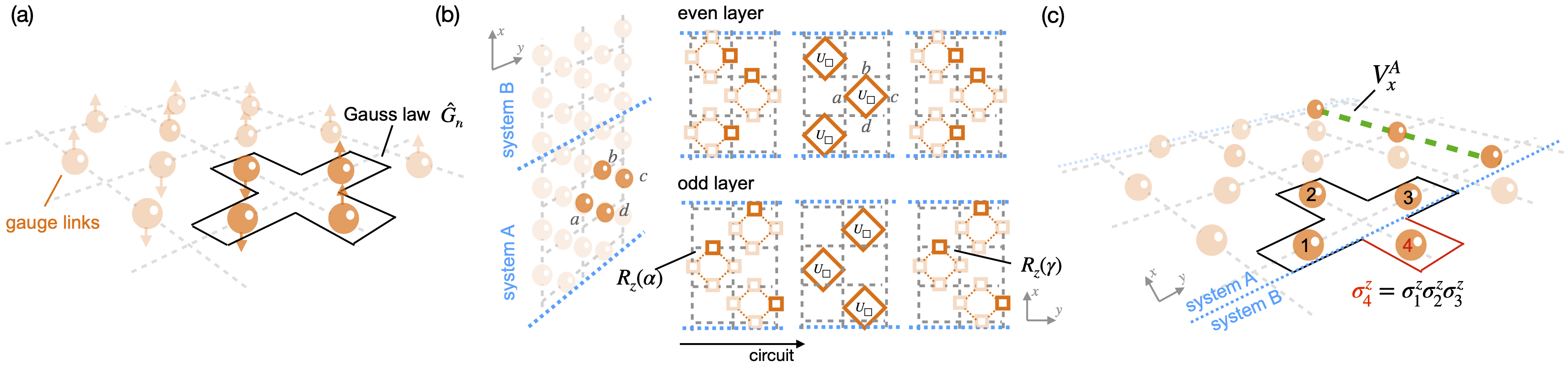}
  \caption{(a) Illustration of  $\Ztwo^{2+1}$ LGT, including Hilbert space and Gauss law constraints $G_j$. (b) Random measurement circuits: We work with an even-odd alternating layers consisting of randomly placed `electric' rotations $R_z(\alpha)$, $R_z(\gamma)$ and plaquette rotations $U_\square(\beta)$ (orange squares), approximating $k$- designs for sufficient circuit depth.  (c) The symmetry structure of $\rho_A$ originates from Gauss laws at entanglement boundaries and a non-local `ribbon' operator $V_x^A$ spanning the two entanglement cuts.\label{fig:2d_setup}}
\end{figure*}
Finally, our scheme can facilitate the experimental identification of topological phases, an inherently difficult task because of the non-detectability of such phases by local measurements~\cite{satzinger2021realizing,semeghini2021probing}. Our proposed strategy is to combine symmetry-conscious randomization with classical shadow or entanglement Hamiltonian tomography to measure the entanglement gap of topologically ordered states, drawing on the foundational work by
Li and Haldane~\cite{li2008entanglement}.

For convenience, we will continue utilizing $k$-designs in a large layer limit---much larger than necessary. 
It is important to reiterate that these choices are not integral to our protocol but are adopted for illustration and convenience, because of the simple channel inversion associated with them for classical shadows and for easier comparison across different schemes. Any tomographically complete scheme is suitable for this task, including shallow depth circuits that are more likely to be realistically employed in near-term experiments. 

We focus on a model related to the toric code, used in ~\cite{satzinger2021realizing}, $\Ztwo$ LGT in (2+1)d spacetime dimensions ($\Ztwo^{2+1}$). This model is graphically illustrated in Fig.~\ref{fig:2d_setup}(a), consisting of spin 1/2 degrees of freedom placed on the links $(j,b)$ of a two-dimensional rectangular $N_x\times N_y$  lattice, where $j=(j_x,j_y)$ and $b=x,y$ is the direction of a link. The Hamiltonian is given by
\begin{align}\label{eq:Z22p1Hamilt}
    H= -K \sum_{j}\sigma^x_{j,x}\sigma^x_{j+\hat{x},y}\sigma^x_{j+\hat{y},x}\sigma^x_{j,y} -g \sum_{j,b} \sigma^z_{j,b}\,,
\end{align}
and superselection sectors are determined by Gauss laws ($[H, \hat {G}_j]=0$),
\begin{align}\label{eq:glaw_z2_2d}
    \hat G_j\equiv \prod_{l\in +(j)} \sigma^z_l,
\end{align}
 where the product is over neighboring links to each lattice site $j$, see \Fig{fig:2d_setup}(a) for illustration where $l$ runs over links labelled (1,2,3,4); physical states obey $\hat{G}_j\ket{\psi}=+1\ket{\psi}$. We assume periodic boundary conditions along the $x$-direction, and fixed or periodic boundary conditions (BCs) in the $y$-direction; for fixed BCs, \Eq{eq:glaw_z2_2d} involves three links at the $y$-boundary. 

Our primary result for this model, and, consequently, for the paper, is to demonstrate that symmetry-respecting randomized measurement schemes enable an appealing and experimentally tractable route to verifying topological order. Ultimately, we shall see that by leveraging the approach known as entanglement Hamiltonian tomography (EHT)~\cite{kokail2021entanglement,kokail2021quantum,zache2022entanglement} and an ansatz for the Entanglement Hamiltonian inspired by the Bisognano-Wichmann theorem~\cite{bisognano1975duality,bisognano1976duality}, we can accurately extract much of the Schmidt spectrum of reduced density matrices $\rho_A$ of the ground state in this model, allowing us to detect topological order via the presence of a so-called `entanglement gap' in the spectrum. 

\subsection{Symmetry-respecting randomized circuits}
We are interested in the entanglement properties of $\rho_A\equiv\mathrm{Tr}_{\bar{A}}(\rho)$, where $A$ is a bipartition obtained by separating the lattice along the $x$-direction, with entanglement cuts at $j_x=0$ and $j_x=N_x^A-1$ so that $N_x=N_x^A+N_x^{\bar{A}}$. The boundary is such that system $A$ contains the  $y$-direction links at $j_x=0$ and $j_x=N_x^A-1$, see \Fig{fig:2d_setup}(c) or \cite{mueller2022thermalization}
for more details, with
$N_\square=(N_x^A-1)\times N_y$ plaquettes ($2N_x^AN_y - N_y - N_x^A$ qubits) in $A$. 
We seek a gauge invariant, symmetry-respecting family of random circuits that form $k$-designs acting on every symmetry sector of $\rho_A$. 
Our ansatz is illustrated in \Fig{fig:2d_setup}(b), made of alternating even-odd half-layers, consisting first of $R_z(\alpha)\equiv \exp\{ i \alpha \sigma^z_i\}$
rotations randomly placed at one side $ i \in \{ a,b,c,d\}$ of a plaquette (orange squares), followed by $U_\square \equiv U_{\square}(\beta) \equiv \exp\{ i \beta \sigma^x_a \sigma^x_b \sigma^x_c\sigma^x_d \} $, and again by a  $R_z(\gamma)\equiv \exp\{ i \gamma \sigma^z_i\}$, placed at the same random $ i \in \{ a,b,c,d\}$. For every plaquette, the angles $\alpha,\beta,\gamma$ are drawn according to a $ZXZ$ decomposition of a CUE matrix,
\begin{equation}
    U_{\rm CUE}\equiv e^{i\delta}\begin{pmatrix}
    e^{i(\alpha+\gamma)}\cos\beta & ie^{-i(\alpha-\gamma)}\sin\beta\\
    ie^{i(\alpha-\gamma)}\sin\beta & e^{-i(\alpha+\gamma)}\cos\beta\\
    \end{pmatrix},
\end{equation}
where the phase $\delta$ is irrelevant. In Appendix~\ref{app:z22ddetails}, we verify numerically that these random circuits form an approximate unitary 2-design ($k$-design).

Before continuing, we discuss the symmetries of $\rho_A\equiv \text{Tr}_{\bar{A}}[\rho]$, depicted in Fig.~\ref{fig:2d_setup}(c).   The Gauss laws at entanglement boundaries allow us to write
\begin{align}\label{eq:symmop2d}
    \sigma^z_4 = \sigma^z_1 \sigma^z_2\sigma^z_3\,.
\end{align}
Here, $\sigma^z_4$ is the `electric field' operator just outside A, and $\sigma^z_1 \sigma^z_2\sigma^z_3$ is just inside A. For all $2N_y$ boundary sites, the operator $\sigma^z_1 \sigma^z_2\sigma^z_3$ is a symmetry of $\rho_A$, i.e. $[\sigma^z_1 \sigma^z_2\sigma^z_3,\rho_A]=0$, if $\rho$ is physical (i.e. Gauss law respecting). We label simultaneous eigensectors of all \Eq{eq:symmop2d} as $s\in\{\ua,\da\}^{2^{2N_y}}$; an example  is $s=\ua\ua\,  \da\ua$ (for $N_y=2$) where the first $N_y$ bits are the eigenvalues of \Eq{eq:symmop2d} at $j_x=0$ and the other $N_y$ at $j_x=N_x^A-1$. Additionally, a `ribbon' operator 
\begin{align}
    V^A_x\equiv \prod_{l \in \mathcal{C}} \sigma^z_l
\end{align}
 commutes with $\rho_A$, $[\rho_A,V^A_x]=0$, where
 $l\in \mathcal{C}$ indicates the links intersected by a contour $\mathcal{C}$ through the centers of plaquettes, from one boundary to the other in an arbitrary path, see the green dashed line in \Fig{fig:2d_setup}(c). For fixed $y$-BCs $V^A_x$ is not  independent  (it is determined by fixing all sectors \Eq{eq:symmop2d}), but for $y$-PBC its eigensectors are independent and labelled by an additional $\ua/\da$, so that $s\in \{ \ua,\da\}^{2^{2N_y+1}}$.
   \begin{figure*}[t]
  \centering
  \includegraphics[scale=0.44]{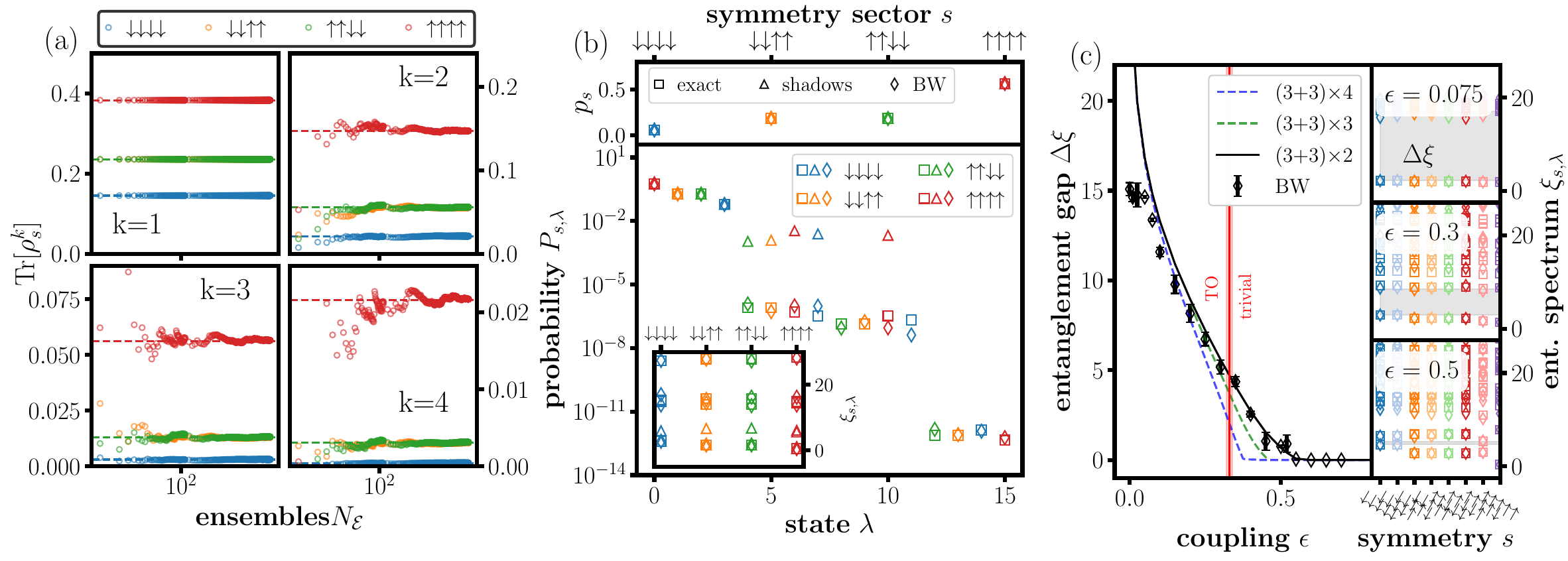}
  \caption{(a) Sector-wise k-purities of $\rho_A$ of the $\mathbb{Z}_{2}^{2+1}$ ground state, from \Eq{eq:kpurities}, for $N_x\times N_y = (3+5)\times 2$,  $\epsilon $=0.1 and $\ell$=64 layers and fixed BC in $y$. (b) Bottom: Shadow- ($N_{\mathcal{S}}=2^{16}$) versus BW-EHT-reconstructed ($N_{\mathcal{E}}=50$, $N_{M}=1024$)  Schmidt spectrum $P_{s,\lambda}$; with $N_x\times N_y = (3+5)\times 2$, $\epsilon $=0.2 and $\ell$=64. Top: probability per sector $p_s$; inset: entanglement spectrum $\xi_{s,\lambda}=-\log(P_{s,\lambda})$. (c) Entanglement gap $\Delta_{\xi}$ between  `low- and high-energy' parts of the ES as a function of $\epsilon$, reconstructed using the BW-EHT scheme, for $N_x\times N_y = (3+3)\times 2$ (PBC in $y$), $N_{\rm BW}=50$, $N_{\rm shots}=1024$, and $\ell=64$. A horizontal red line is the infinite volume limit, $\epsilon_c=0.33\pm0.01$. Right: Symmetry-resolved Entanglement spectra $\xi_{s,\lambda}$ for $\epsilon=0.075,03,0.5$.\label{fig:LGT2dEHT}}
\end{figure*}

\subsection{Classical shadows}
As a warm-up, we first study again $k$-purities for the new model, thereby verifying correctness of the randomization, before moving on to reconstructing a classical shadow representation of the state that we will use to infer the presence or absence of TO.  In Fig.~\ref{fig:LGT2dEHT}(a) we display  $k$-purities $\mathrm{Tr}[\rho_{A,s}^k]$ ($k=1,2,3,4$) of the $\mathbb{Z}_2^{2+1}$ ground state at $\epsilon\equiv g/K=0.1$, following the approach outlined in Sect. \ref{s:gaugetheories}, as a function of $N_{\mathcal{E}}$ and in the
infinite shot limit  ($N_M\rightarrow\infty$), with fixed $y$-BCs,  $N_x\times N_y = (3+5)\times 2$ and $\ell=64 $. As before, the $k=1$ results, $p_s =\text{Tr}[\rho_{A,s}]$, are exact by design while $k=2,3,4$-entropies are reproduced with increasing  $N_{\mathcal{E}}$. 

Finally, aiming at reconstructing the entanglement spectrum from classical shadows to diagnose TO, in \Fig{fig:LGT2dEHT}(b), we  show the Schmidt spectrum $P_{s,\lambda}$ for $\epsilon=0.2$, comparing shadow-reconstructed ($N_{\mathcal{S}}=2^{16}=65,536$, triangles) versus exact results (squares). Different colors represent symmetry sectors $s$ with weights $p_{s} \equiv \text{Tr}_s [\rho_{A,s}] $  in the top panel. Large Schmidt values are well reproduced  down to approximately $10^{-2}$, beyond which we observe significant deviations that prevent us from reliably determining the entanglement gap. 

\subsection{Detecting topological order through entanglement Hamiltonian tomography}
Aiming at higher precision, we explore an alternative approach, Entanglement Hamiltonian Tomography (EHT)~\cite{kokail2021entanglement,kokail2021quantum,zache2022entanglement}. The basic idea is to parameterize the reduced density matrix by an Entanglement Hamiltonian (EH), 
\begin{align}
    H_A=-\log{ [\rho_A]}\,.
\end{align}
A (heuristic) parameterization of ground state EHs, inspired by the Bisognano-Wichmann (BW) theorem~\cite{bisognano1975duality,bisognano1976duality}, is 
\begin{align}\label{eq:BW}
    H_A \equiv H_A[\{\beta_{\mathcal{O}} \}] = \sum_{\mathcal{O}} \beta_{\mathcal{O}} H_{\mathcal{O}}\,,
\end{align}
where $ H_{\mathcal{O}}$ are the local operators comprising the physical Hamiltonian, \Eq{eq:Z22p1Hamilt}, and $\beta_{\mathcal{O}}$ a local `temperature', varying with the distance of $\mathcal{O}$ from the entanglement cut(s) (also depending on $j_y$ if translation invariance in $y$ is broken, i.e. for fixed $y$-BCs). The applicability and accuracy of this ansatz was investigated for $\mathbb{Z}_2^{2+1}$ in~\cite{mueller2022thermalization}.

To extract the $\{\beta_{\mathcal{O}} \}$, we follow~\cite{kokail2021entanglement} and first measure probabilities $P_U(b,s)$ in $N_{\mathcal{E}}$ random bases and with $N_M$ shots each. We then minimize 
\begin{align}
    \sum_{b} \Big\langle \Big(  P_U(b,s) - \text{Tr}_s \big[  \bar{\rho}_{A,s}U_s | b,s \rangle \langle b,s | U^\dagger_s  \big] \Big)^2\Big \rangle_{ \mathcal{E}},
\end{align}
with respect to $\{\beta_i\}$ via classical post-processing, where $ \bar{\rho}_{A,s}\equiv \bar{\rho}_{A,s}[\{\beta_{\mathcal{O}} \} ] \sim\exp\{ -H_{A,s}[\{\beta_{\mathcal{O}} \} ]\}$, normalized so that $\text{Tr}_s[\bar{\rho}_{A,s}[\{\beta_{\mathcal{O}} \} ]=p_s$. $H_{A,s}[\{\beta_{\mathcal{O}} \} ]\equiv \sum_{\mathcal{O}}\beta_{\mathcal{O}} H_{s,\mathcal{O}}$ is the EH with $H_{s,\mathcal{O}}$ restricted to a symmetry sector $s$. Because the BW optimization is performed sector-wise, matrices of size $d_s$ are involved, versus  the dimension of $A$, $d_{A}$. In the infinite measurement limit,  the optimization will yield one universal set $\{\beta_{\mathcal{O}}\}$ for all $s$, but in practice we work with normalized $\bar{\rho}_{A,s}\equiv {\rho}_{A,s}/p_s$ and $\bar{P}_U(b,s) = {P}_U(b,s) / p_s$, so that the extracted $\{\beta_{\mathcal{O},s}\}$ depend on that normalization and differ from $\{\beta_{\mathcal{O}}\}$,  see Appendix \ref{app:z22ddetails} for details.

Results of the BW-EHT-optimization for the Schmidt spectrum $P_{s,\lambda}$ are displayed as diamond symbols in \Fig{fig:LGT2dEHT}(b) along with shadow results, for $N_{\mathcal{E}}=50$  and $N_M=1024$. Despite comparable cost ($N_{\mathcal{E}}\cdot N_M=51,200$) relative to the classical shadow approach ($N_{\mathcal{S}}=2^{16}=65,536$),  BW-EHT  reproduces the eigenvalue spectrum much more accurately; values as small as $10^{-6}$ are approximately recovered and even eigenvalues as small as $10^{-11}$ are not far off. The inset of \Fig{fig:LGT2dEHT}(b) shows the entanglement spectrum (ES), i.e. the spectrum of the EH, which is also well reproduced.
The apparent advantage of the BW-EHT approach comes at the expense of generality, it is tailored for ground states (it can be extended to non-equilibrium states~\cite{kokail2021entanglement,mueller2022quantum}) while classical shadows work regardless of the state. 

Enabled by the performance of the BW-EHT optimization, we focus on a practical application: detecting topological order (TO) of quantum states. Ground states of $\mathbb{Z}_2^{2+1}$ are separated (in the infinite volume limit) into topologically ordered, $\epsilon < \epsilon_c$, and trivial states, $\epsilon > \epsilon_c$, with a phase transition at a critical coupling $\epsilon_c$. Li and Haldane's entanglement-boundary conjecture ~\cite{li2008entanglement,mueller2022thermalization}
asserts that TO states are `entanglement-gapped', i.e. their ES has separated low energy (large Schmidt values)
and a high energy (small Schmidt values) parts. Further, the low lying part is (up to rescaling) identical to the  spectrum of a conformal field theory (CFT) describing gapless excitations at the edge of the system. We focus here on measuring the existence of an entanglement gap $\Delta\xi$ to detect TO, which has been shown as very a robust order parameter for the TO transition in this model even for very small systems~\cite{mueller2022thermalization}.

Without loss of generality, to reduce finite size effects, we focus on periodic boundary conditions in $y$ (a torus) for the BW-EHT analysis. In \Fig{fig:LGT2dEHT}(c) we show the entanglement gap $\Delta \xi$ for $N_x \times N_y = (3+3)\times 2$,  $N_{\mathcal{E}}=50$, $N_M=1024$, and $\ell=64$ (black diamonds), compared to exact results (black solid line). Error bars represent the combined statistical error due to finite $N_{\mathcal{E}}$ and $N_M$, see Appendix \ref{app:z22ddetails} for details. A vertical red line indicates the infinite volume extrapolated value $\epsilon_c = 0.33\pm 0.01$~\cite{blote2002cluster}. We also show result for $N_x \times N_y = (3+3)\times 3$ (green dashed line) and $N_x \times N_y = (3+3)\times 4$ (blue dashed line), taken from~\cite{mueller2022thermalization} and approaching the infinite volume limit to within less than 10\%. Side panels show the BW-EHT-reconstructed sector-wise ES for $\epsilon=0.075,0.3,0.5$, demonstrating the closing of the entanglement gap (gray shaded area) at $\epsilon_c$ where our results approximately reproduce the phase transition.

We could not numerically simulate systems larger than $N_x \times N_y=(3+3)\times 2$. For example, a $N_x \times N_y=(3+3)\times 4$ lattice of 48 qubits (20 qubits in the subsystem) exhausts our classical computational resources.\footnote{The results for $N_x \times N_y=(3+3)\times 4$ and $(3+3)\times3$ were obtained using exact diagonalization and working with dual formulations of $\mathbb{Z}_2^{2+1}$~\cite{mueller2022thermalization}.} However,  the classical (shadow- or BW-EHT-) analysis is simple for such a system if it were prepared in experiment because, while $d_{A}=2^{2N_x^A N_y-N_y}=2^{20}=1,048,576$, the analysis is restricted to symmetry blocks of only $d_s=2^{N_x^A N_y-N_y}=2^{8}=256$ states. This is a significant (in fact, exponential) reduction in the space over which the state is randomized, but $d_s$ still grows exponentially with the subsystem size, albeit much slower than $d_A$. 

\section{Conclusion and Outlook}\label{s:conclusion}
In this manuscript, we proposed randomized measurement protocols for lattice models that leverage symmetries, focusing primarily on LGT entanglement structure exploration. We devised deep-scrambling circuits that realize symmetry-conscious $k$-designs and illustrated their use in simple gauge and non-gauge model examples. Our approach is intuitive and, therefore, easily generalizable: by examining the physical Hamiltonian one can readily identify basic symmetry-preserving interactions which can be used as the generators of a randomized measurement scheme. Consequently, if a particular physical Hamiltonian can be realized, so can our measurement scheme. 

Symmetry-conscious randomized measurement schemes like those considered here have lower sampling costs compared to symmetry-ignorant schemes by avoiding randomizing over non-relevant Hilbert space parts.  In particular, one obtains a sampling cost (to realize a $2$- ($k$-)design) that scales with block size $d_s$, instead of  Hilbert space dimension $d_{\mathcal{H}}$. This reduction can be exponential, e.g., for particle number conserving systems away from half-filling, or for LGTs due to randomizing only over the physical sector of Hilbert space. In constructing $k$-designs, while still efficient, they also incur a somewhat larger circuit complexity.

Using such symmetry-conscious randomized measurement, our primary goal was to provide a practical scheme for measuring LGT entanglement structure, a potential useful route e.g.,
for quantum simulating high energy and nuclear physics~\cite{carlson2018quantum,cloet2019opportunities,beck2019nuclear,davoudi2022quantum,catterall2022report,beck2023quantum}, e.g. to understand Quantum Chromodynamics (QCD) where entanglement is largely unexplored~\cite{kharzeev2022quantum,cervera2017maximal,beane2019entanglement,beane2021geometry,beane2021entanglement,klco2021geometric,klco2021entanglement,klco2021entanglementspheres,mueller2022thermalization}, or detecting topologial order.
We illustrated our approach in a simple (1+1)d LGT example, $\mathbb{Z}_2$ coupled to staggered matter, where we extracted symmetry-resolved $k$-purities and von Neumann entanglement entropies, and separated their symmetry- and distillable components. 

We then focused on $\mathbb{Z}_2$ in $2+1$ dimensions where the intricate structure of gauge symmetric states can lead to topologically ordered (TO) phases. These are currently receiving great attention, including experimental  realizations in AMO and solid-state platforms~\cite{satzinger2021realizing,semeghini2021probing}, motivated by applications such as fractional quantum Hall effect states~\cite{stormer1999fractional,cage2012quantum} or fault-tolerant quantum computation and storage~\cite{c1982stormer,wen1990topological,kitaev2003fault,kitaev2006anyons,sarma2006topological,nayak2008non,sarma2015majorana,lahtinen2017short}. A difficulty is that TO cannot be probed by measuring local operators, a serious impediment for its experimental verification. Our approach to overcome this is based on measuring the entanglement structure of such systems. While the importance of entanglement as a robust indicator of topological order was realized long ago~\cite{levin2005string,levin2006detecting,kitaev2006topological}, we developed a concrete random-measurement scheme, following the logic of Li and Haldane~\cite{li2008entanglement}, that uses a state presentation in terms of Entanglement Hamiltonians (EH) and is based on measuring entanglement gaps of their (symmetry-resolved) spectrum using a tomographic protocol based on the Bisognano-Wichmann theorem~\cite{kokail2021entanglement}. Remarkably, performing random measurements on very small  subsystems as small as $N_x^A \times N_y = 3\times 2$, we observe a relatively sharp TO-to-trivial phase transition. While our focus was on $\mathbb{Z}_2^{2+1}$, the protocol 
can be easily generalized to other systems. 

A benefit of our approach, not explicitly explored in the main text, is that symmetry-conscious randomization allows for a rudimentary, but useful, near-term error-mitigation strategy similar to that discussed in~\cite{nguyen2022digital}. A feature of symmetry-conscious randomization is that symmetries of the input states are not lost and can be measured. Thus, machine errors that violate those symmetries are detectable after randomization, suggesting that e.g., a postselection of measurement results can improve the computation (at the cost of reduced statistics). Another potential application is approximate Haar random state preparation for thermal state algorithms~\cite{davoudi2022toward}. 
Finally, we expect symmetry-respecting randomized measurement schemes to be useful to investigate thermal systems, including e.g, systems with non-Abelian conserved charges~\cite{guryanova2016thermodynamics, yunger2016microcanonical, lostaglio2017thermodynamic, halpern2018beyond,halpern2020noncommuting, fukai2020non, popescu2020reference, yunger2022build, kranzl2022experimental, manzano2022non,  mitsuhashi2022char, majidy2023non}.

There are many future extensions of our work. For example, while our approach significantly reduces algorithmic costs compared to a symmetry-ignorant scheme, extracting entanglement entropies and structure still relies on classical post-processing which ultimately scales exponentially with system size, an issue which can be addressed with quantum variational~\cite{kokail2021quantum} and machine learning techniques~\cite{huang2022learning,huang2022quantum}. Extending the robust numerical analysis performed here to provide analytical performance guarantees for the circuit depth and sampling complexity of the random circuits in this work is also of clear interest. We also emphasize that realizing an approximate $k$-design is a sufficient, but not necessary, condition for randomized measurement protocols, and more studies regarding optimal randomization for certain observables are needed.\footnote{During the review process of this manuscript, some interesting analytical results along these lines for circuits with particle number symmetry were presented in~\cite{hearth2023unitary}, see also~\cite{hearth2023efficient} for shallow depth circuits.} 
Developing a formalism for shadow and entanglement tomography protocols that applies to systems with limited control or is independent of the circuit model would be useful for analog quantum simulation.
Finally, while there are encouraging indications~\cite{senrui2021robust, koh2022classicalshadows, tran2022measuring}, the robustness of our scheme against experimental imperfections and noise should be investigated in detail.

Finally, we point out related work~\cite{zhao2021fermionic, hao2022classical}, following a similar idea for fermionic systems with particle number symmetry, and also demonstrating a significant cost advantage. While not programmable enough to realize $k$-designs, we think this is a very useful approach for fermionic entanglement tomography. 
We also point out Ref.~\cite{van2022hardware}, proposing randomized measurement schemes that take advantage of \emph{a priori} knowledge about  observables of interest to improve sampling complexity and  Refs.~\cite{vittorio2022symmetry,rath2023entanglement}, which, building on Ref.~\cite{neven2021symmetry}, propose randomized measurement schemes to extract symmetry-resolved purities. Their scheme is based on local random unitary transformations, in contrast to the symmetry-preserving unitaries we consider here.

\section*{Acknowledgments}
We thank Alexander F. Shaw for early discussions leading to the formulation of this project.
We also thank 
Zohreh Davoudi, 
Jonas Helsen, Hsin-Yuan (Robert) Huang, 
Martin Savage, Guilia Semeghini, Nicole Yunger Halpern, Torsten Zache and Peter Zoller for discussions.
J.B. was funded in part by the Heising-Simons Foundation, the Simons Foundation, and National Science Foundation Grant No. NSF PHY-1748958 and by the U.S. Department of Energy (DOE) ASCR Accelerated Research in Quantum Computing program (award No.~DE-SC0020312), DoE QSA, NSF QLCI (award No.~OMA-2120757), DoE ASCR Quantum Testbed Pathfinder program (award No.~DE-SC0019040), NSF PFCQC program, AFOSR, ARO MURI, AFOSR MURI, and DARPA SAVaNT ADVENT. N.M. acknowledges funding by the U.S. Department of Energy, Office of Science, Office of Nuclear Physics, InQubator for Quantum Simulation (IQuS) (\url{https://iqus.uw.edu}) under Award Number DOE (NP) Award DE-SC0020970; and, during early stages, by the U.S. Department of Energy’s Office of Science, Office of Nuclear Physics under Award no. DE-SC0021143 for quantum simulation of gauge-theory dynamics on near-term quantum hardware. This work was enabled, in part, by the use of advanced computational, storage and networking infrastructure provided by the Hyak supercomputer system at the University of Washington~\cite{hyak}.

\appendix

\section{Equivalence of error metrics}\label{app:equiv_error_metrics}
Our choice of error metric is the average difference between matrix elements of $\mathcal{B}^s$, Eq.~(\ref{eq:2designquant}), calculated from an ensemble of unitaries generated by random circuits and those obtained from an ensemble of unitaries that form an exact 2-design, as given in Eq.~(\ref{eq:def2design}). This difference is normalized by a factor of $d_s(d_s^2-1)$ where $d_s$ is the dimension of the relevant symmetry block. Because of the otherwise prohibitive cost, in practice, for large $N$ we compute the error by only averaging over \emph{a sample} of all matrix indices of $\mathcal{B}^s$, where we increase the number of indices sampled until we see convergence. Mathematically,
\begin{equation}\label{eq:ourerror}
\epsilon = \frac{d_s(d_s^2-1)}{|S|}\sum_S \big|(\mathcal{B}_\mathcal{E}^s)^{ i'j'k'l'}_{ijkl}-(\mathcal{B}^s_{2\text{-des.}})^{ i'j'k'l'}_{ijkl}\big|,
\end{equation}
where $S$ is the set of indices $i, j, k, l, i', j', k' , l'$ sampled and the subscripts on $\mathcal{B}^s$ denote the ensemble with which the expectation values are taken with respect to in the definition of $\mathcal{B}^s$.

This choice of quantifying the error between our random ensemble of circuits and an exact unitary $2$-design was chosen for two reasons: (1) it enables a computationally tractable approach wherein we sample a subset of matrix elements until we see convergence; (2) once normalized (by a factor of $d_s(d_s^2-1)$), it is manifestly dimension independent once our random ensemble of circuits has converged to an approximate unitary $2$-design. Despite these benefits, it is not the only reasonable choice of error metric, nor is it a standard one. Therefore, in this appendix, we demonstrate analytically that it is equivalent to more typical definitions, up to a rescaling of the approximation ratio for the case where $S$ is the set of all index combinations. We then numerically demonstrate that sampling (the non-zero) matrix elements is also a valid (and computationally accessible) choice of error metric. 

A particularly common definition of an approximate unitary $k$-design states that an ensemble $\mathcal{E}$ of unitaries forms a $\delta$-approximate unitary $k$-design if and only if~\cite{hunter2019unitary}
\begin{equation}\label{eq:frame-pot-def}
\sqrt{|\mathcal{F}^{(k)}_\mathcal{E}-\mathcal{F}_\mathrm{Haar}^{(k)}
|}\leq \frac{\delta}{d_s^k},
\end{equation}
where $\mathcal{F}^{(k)}$ is the $k$-th frame potential for an ensemble of unitaries, defined as
\begin{equation}
\mathcal{F}_\mathcal{E}^{(k)}=\int_{U,V\in\mathcal{E}}dU dV\left|\mathrm{Tr}(U^\dagger V)\right|^{2k}.
\end{equation}
For the Haar ensemble $\mathcal{F}^{(k)}_\mathrm{Haar}=k!$ for $k\leq d_s$~\cite{gross2007evenly}. 

Ultimately, we will numerically compare our error metric to the frame potential definition, but for the purposes of analytically showing equivalence of our error metric it is convenient to use a different formulation known to be equivalent to this one up to a rescaling of the approximation factor~\cite{low2010pseudo}. In particular, an ensemble $\mathcal{E}$ of unitaries forms a $\delta$-approximate $k$-design if and only if for all balanced monomials $M(U)$ of degree $\leq k$ in the matrix elements of $U$
\begin{equation}\label{eq:defkdesign}
\big |\langle M(U)\rangle_\mathcal{E}-\langle M(U)\rangle_{k} \big|\leq \frac{\delta}{d_s^k},
\end{equation}
where the subscript $k$ denotes an expectation value with respect to an exact $k$-design. A balanced monomial of degree $k$ in the matrix elements of $U$ is defined as any product in the matrix elements of the form
\begin{equation}
M=U_{i_1 j_1}...U_{i_k j_k}U_{k_1 l_1}^*...U_{k_k l_k}^*,
\end{equation}
for some choice of indices.

To show the equivalence between Eq.~(\ref{eq:ourerror}) and Eq.~(\ref{eq:defkdesign}) we return explicitly to the case of $k=2$. An ensemble $\mathcal{E}$ that forms a $\delta$-approximate $2$-design necessarily also forms a $\delta$-approximate $1$-design, so in Eq.~(\ref{eq:defkdesign}) we can restrict our attention to balanced monomials of degree 2. 
Consequently, for unitaries $U^s$ acting on a symmetry sector $s$, and a general degree 2 balanced monomial $M=U^s_{ij} U^{s*}_{i'j'}U^s_{kl} U^{s*}_{k'l'}$, we can write
\begin{widetext}
\begin{align}\label{eq:a1}
&\big|(\mathcal{B}_\mathcal{E}^s)^{ i'j'k'l'}_{ijkl}-(\mathcal{B}^s_{2})^{ i'j'k'l'}_{ijkl}\big|= \nonumber \\
&\bigg |\langle M_{ijkl}^{ i'j'k'l'}\rangle_{\mathcal{E}}
     -\langle M_{ijkl}^{ i'j'k'l'}\rangle_{2}+ \frac{d_s^2}{d_s^2-1} \Big[ (\mathcal{A}_{2}^s)^{i'j'}_{ij}(\mathcal{A}_{2}^s)^{k'l'}_{kl}+ (\mathcal{A}_{2}^s)^{k'l'}_{ij}(\mathcal{A}_{2}^s)^{i'j'}_{kl}-(\mathcal{A}_{\mathcal{E}}^s)^{i'j'}_{ij}(\mathcal{A}_{\mathcal{E}}^s)^{k'l'}_{kl}-(\mathcal{A}_{\mathcal{E}}^s)^{k'l'}_{ij}(\mathcal{A}_{\mathcal{E}}^s)^{i'j'}_{kl}\Big]\bigg | \nonumber \\
     & \leq \underbrace{\bigg |\langle M_{ijkl}^{ i'j'k'l'}\rangle_{\mathcal{E}}
     -\langle M_{ijkl}^{ i'j'k'l'}\rangle_{2}\bigg|}_{\text{I}}+ \frac{d_s^2}{d_s^2-1} \Bigg[\underbrace{\Big| (\mathcal{A}_{2}^s)^{i'j'}_{ij}(\mathcal{A}_{2}^s)^{k'l'}_{kl}-(\mathcal{A}_{\mathcal{E}}^s)^{i'j'}_{ij}(\mathcal{A}_{\mathcal{E}}^s)^{k'l'}_{kl}
     \Big |}_{\text{II}} 
     +
     \underbrace{\Big|(\mathcal{A}_{2}^s)^{k'l'}_{ij}(\mathcal{A}_{2}^s)^{i'j'}_{kl}-(\mathcal{A}_{\mathcal{E}}^s)^{k'l'}_{ij}(\mathcal{A}_{\mathcal{E}}^s)^{i'j'}_{kl}\Big|}_{\text{II}^*}\Bigg]
\end{align}
where the matrix elements $(\mathcal{B}^s_{2})^{i'j'k'l'}_{ijkl}$ are the matrix elements for an exact unitary 2-design, as defined in  Eq.~(\ref{eq:def2design}) and, as noted in the main text, $(\mathcal{A}^s)_{ij}^{kl}\equiv\langle U^s_{ij}U^{s*}_{kl} \rangle$. The subscript $2$ denotes an average taken with respect to an exact unitary 2-design and the subscript $\mathcal{E}$ denotes an average taken with respect to the corresponding ensemble of unitaries. 

For a $\delta$-approximate unitary $k$-design the term labeled I is bounded via Eq.~(\ref{eq:defkdesign}). The terms labeled II and II$^*$ are equivalent up to the choice of indices. Recalling that for an exact 1-design (and thus also for a 2-design) $(\mathcal{A}^s_{2})_{ij}^{kl}=(\mathcal{A}^s_{1})_{ij}^{kl}=\frac{\delta_{ij}\delta_{kl}}{d_s}$ and applying Eq.~(\ref{eq:defkdesign}) for an approximate unitary $1$-design, we can bound these terms as
\begin{align}
&\text{II} = \Big| (\mathcal{A}_{2}^s)^{i'j'}_{ij}\Big ((\mathcal{A}_{\mathcal{E}}^s)^{k'l'}_{kl}-(\mathcal{A}_{2}^s)^{k'l'}_{kl}\Big)+ (\mathcal{A}_{2}^s)^{k'l'}_{kl}\Big ((\mathcal{A}_{\mathcal{E}}^s)^{i'j'}_{ij}-(\mathcal{A}_{2}^s)^{i'j'}_{ij}\Big) +
\Big ((\mathcal{A}_{\mathcal{E}}^s)^{i'j'}_{ij}-(\mathcal{A}_{2}^s)^{i'j'}_{ij}\Big)\Big ((\mathcal{A}_{\mathcal{E}}^s)^{k'l'}_{kl}-(\mathcal{A}_{2}^s)^{k'l'}_{kl}\Big)\Big |\nonumber \\
&\leq \Big| (\mathcal{A}_{2}^s)^{i'j'}_{ij}\Big ((\mathcal{A}_{\mathcal{E}}^s)^{k'l'}_{kl}-(\mathcal{A}_{2}^s)^{k'l'}_{kl}\Big)\Big |+ \Big |(\mathcal{A}_{2}^s)^{k'l'}_{kl}\Big ((\mathcal{A}_{\mathcal{E}}^s)^{i'j'}_{ij}-(\mathcal{A}_{2}^s)^{i'j'}_{ij}\Big) \Big |+\Big |
\Big ((\mathcal{A}_{\mathcal{E}}^s)^{i'j'}_{ij}-(\mathcal{A}_{2}^s)^{i'j'}_{ij}\Big)\Big ((\mathcal{A}_{\mathcal{E}}^s)^{k'l'}_{kl}-(\mathcal{A}_{2}^s)^{k'l'}_{kl}\Big)\Big |\nonumber \\
&\leq \frac{\delta}{d_s^2}+\frac{\delta}{d_s^2}+\frac{\delta^2}{d_s^2}.
\end{align}
\end{widetext}

Plugging back into Eq.~(\ref{eq:a1}) we obtain
\begin{align}
\big|(\mathcal{B}_\mathcal{E}^s)^{ i'j'k'l'}_{ijkl}-(\mathcal{B}^s_{2})^{ i'j'k'l'}_{ijkl}\big| &\leq \frac{\delta}{d_s^2}+\frac{2d_s^2}{d_s^2-1}\left(\frac{2\delta+\delta^2}{d_s^2}\right) \nonumber \\
&\hspace{-2em}=\frac{\delta}{d_s^2}\left(\frac{d_s^2(5+2\delta)-1}{d_s^2-1}\right),
\end{align}
for any choice of indices. 
It is straightforward to go through the converse of this argument and show that if $|\mathcal{B}_\mathcal{E}-\mathcal{B}_2|<\frac{\delta}{d_s^2}$ and $|\mathcal{A}_\mathcal{E}-\mathcal{A}_2|<\frac{\delta}{d_s}$ (indices suppressed), then the approximate two-design property for $\langle M\rangle_\mathcal{E}$ holds, up to a rescaling of $\delta$.
Consequently, up to an inconsequential rescaling of the approximation factor, any norm on the difference of the tensors $(\mathcal{B}_\mathcal{E}^s)-(\mathcal{B}_2^s)$ is a valid choice for defining an approximate unitary $2$-design consistent with the definition in Eq.~(\ref{eq:defkdesign}).  

Our error metric, if all indices are sampled, is just such a norm (namely, an element-wise 1-norm, up to dimension-dependent factors). For the sake of computational tractability, however, we randomly sample a collection of the non-zero matrix elements of $\mathcal{B}^s$ until we see convergence. If the sampled average has converged statistically, i.e. has small error bars, this is a good indication that the sampled average accurately reflects the total average, since the total is simply a sum of sample averages. Since the matrix elements of $\mathcal{B}^s_2$ can take only a handful of values and are all either zero or of order $d_s^{-3}$, we expect there are no "outlier" matrix elements whose errors are systematically larger than others. We have explicitly verified that the sampling scheme converges to the results obtained by evaluating all index combinations, including those where Eq.~(\ref{eq:def2design}) yields zero. In all our numerics, we ensure that the number of samples taken is such that this convergence occurs.  

\begin{figure}
    \centering
    \includegraphics[width=0.98\columnwidth]{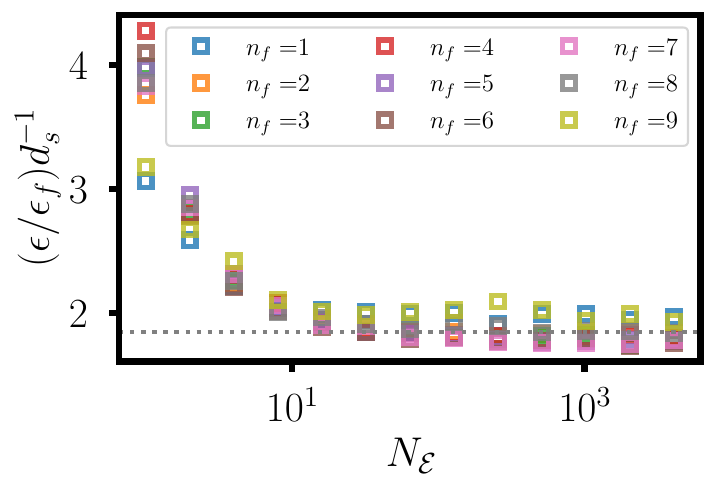}
    \caption{Ratio of our error metric $\epsilon$ [given in \Eq{eq:ourerror}] to the more standard frame potential error metric $\epsilon_f$ [given in \Eq{eq:frame-pot-def}], for an approximate unitary $2$-design per symmetry sector (labeled by particle number $n_f$), normalized by $d_s$, versus number of unitaries $N_\mathcal{E}$ sampled, indicating that the two error metrics are equivalent up to a rescaling of the approximation ratio for sufficiently large $N_\mathcal{E}$. Results are for a system of $N=10$ sites with particle-number symmetry as described in Sect.~\ref{sec:globalsymmetry}. For both error metrics, we sample directly from the CUE within each symmetry sector. Our error metric is computed by averaging over $|S|=2000$ non-zero matrix elements.}
    \label{fig:compare-error-metrics}
\end{figure}

Numerical results demonstrating the equivalence of our error metric when using finite samples of the non-zero matrix elements of $\mathcal{B}^s$, as given in \Eq{eq:ourerror}, to the error metric in terms of the frame potential in Eq.~(\ref{eq:frame-pot-def}) are shown in \Fig{fig:compare-error-metrics}. Here, we consider a system of $N=10$ sites with particle number symmetry as presented in Sect.~\ref{sec:globalsymmetry} of the main text. For this comparison, we compare the two error metrics as a function of the number of unitaries $N_\mathcal{E}$ drawn directly from CUE within each symmetry sector. In the next section, we elaborate on how we demonstrated that the random circuits we describe in the main text do, in fact, converge to sampling from the CUE for a sufficient number of layers. However, temporarily leaving this aspect aside allows us to purely compare the two error metrics. We find that the two error metrics are equivalent up to a rescaling of the approximation ratio by a factor $\sim d_s$ for a sufficiently large ensemble of random unitaries. This is consistent with, but tighter, than our analytic results. Also consistent with our analytics, the numerics indicate sub-leading $d_s$-dependent factors in the rescaling of the error.

\section{Particle number analysis}\label{app:numerical_details}
In this appendix, we provide additional details for the particle number symmetry analysis, presented in Sect.~\ref{sec:globalsymmetry} of the main text. In particular, we present the numerical evidence that sufficiently deep random symmetry-respecting circuits of the sort we describe there form an approximate $2$-design and provide details on the numerics leading to Fig.~\ref{fig:scalingpn}.

For both of these purposes, we compute the error $\epsilon$ as defined in Eq.~(\ref{eq:ourerror}) for ensembles $\mathcal{E}$ of $N_\mathcal{E}$ random unitaries generated both by our random circuits and by drawing directly from the CUE. Note that in the latter case, the error is purely due to the fact that these ensembles have only a finite number of elements $N_\mathcal{E}$, as for $N_\mathcal{E}\rightarrow\infty$ the ensemble will be a $2$-design by construction. Consequently, as can be seen in Fig.~\ref{fig:scalingvslayers}a, the error $\epsilon$ scales as $\sim 1/\sqrt{N_\mathcal{E}}$.

For a sufficient number of layers, the ensembles generated by our random circuits have identical error $\epsilon$ as the ensembles drawn directly from CUE, indicating that these circuits indeed generate samples that form an approximate $2$-design. This is demonstrated in Fig.~\ref{fig:scalingvslayers}b, where we plot the difference between the error $\epsilon$ obtained by taking $N_\mathcal{E}=8192$ samples from the random circuits and taking $N_\mathcal{E}=8192$ samples directly from the CUE versus number of layers used in the circuits. Within a short depth $\ell\approx 15$, the difference reaches a floor set by $N_\mathcal{E}$ (i.e. the standard $1/\sqrt{N_\mathcal{E}}$ sampling error), indicating that at this depth, drawing samples from our circuits is equivalent to drawing samples directly from CUE. All of the remaining error in approximating a $2$-design is purely due to the finite number of samples $N_\mathcal{E}$. 

While the saturation point (in terms of number of layers required) in Fig.~\ref{fig:scalingvslayers}(b) shows a slight dependence on the dimension $d_s$ and similar numerics for smaller system sizes shows a dependence on the number of sites $N$, the fact that our numerics are limited to around $N
\le 10$ sites prevents us from extracting an asymptotic scaling form of these dependencies. For the pragmatic approach taken in this work---namely, demonstrating that our approach is a viable one for extracting quantities of interest for specific systems with symmetries of interest, such as $\mathbb{Z}_2$ LGT, for small to moderate system sizes---these numerics are sufficient. However, a detailed analytic analysis of the scaling of the error with the number of layers for the particular models considered here (or, perhaps, generally for random symmetry-respecting circuits), remains a compelling prospect for future work. Such analyses have been done for symmetry-ignorant designs~\cite{dulian2022random} and similar approaches should apply here~\cite{hearth2023efficient}.

\subsection{Details for Figure 3(a)}
\begin{figure}
    \centering
    \includegraphics[width=0.98\columnwidth]{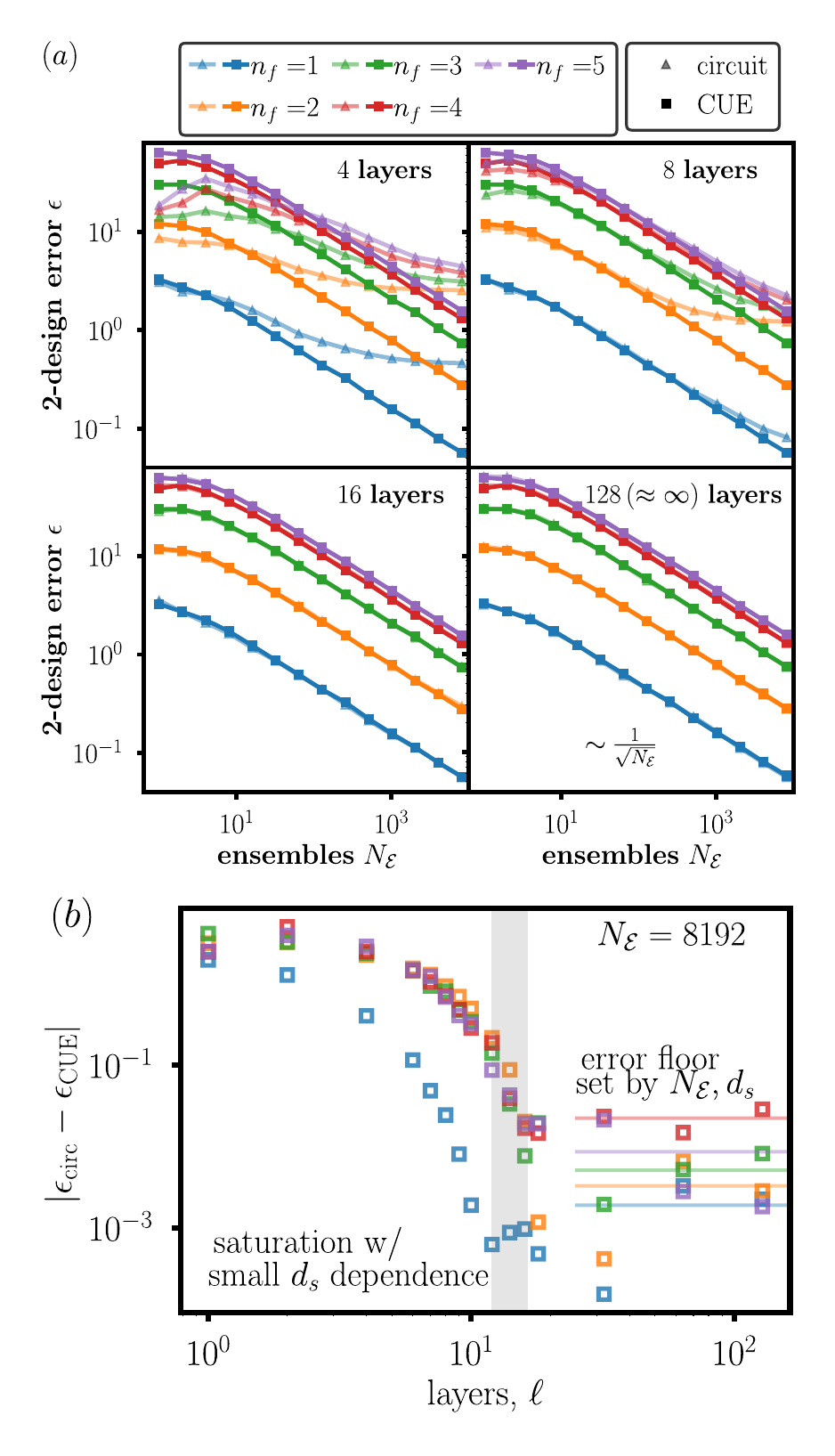}
    \caption{(a) Error relative to an exact unitary $2$-design $\epsilon$ (Eq.~(\ref{eq:def2design})) per particle number symmetry sector (labeled by $n_f$) versus number of unitaries $N_\mathcal{E}$. For clarity only particle number sectors $n_f\leq 5$ are shown; as they are of equivalent dimension, particle number sectors for $n_f>5$ have the same behavior as the particle number sector $(10-n_f)$. The ensemble $\mathcal{E}$ is sampled both from the particle number symmetry-respecting random circuits with different numbers of layers $\ell\in\{4,8,16,128\}$ (triangles) and sampled directly from the CUE within each symmetry sector (squares). Results shown are for $n=10$ sites and the error is averaged over 2000 non-zero matrix elements. Extrapolation of fits to the error as a function of $N_\mathcal{E}$ at $\ell=128$ layers for the data pictured here and equivalent data for $N=4,6,8$ sites is used to produce \Fig{fig:scalingpn}a of the main text. (b) The difference between the error $\epsilon$ obtained by taking $N_\mathcal{E}=8192$ samples from the random circuits and taking $N_\mathcal{E}=8192$ samples directly from the CUE versus number of layers used in the circuits. By $\ell\approx 15$, the difference reaches a floor set by $N_\mathcal{E}$ (i.e. the standard sampling error) with a small dependence on the dimension $d_s$ of the corresponding sector.}
    \label{fig:scalingvslayers}
\end{figure}

Fig.~\ref{fig:scalingvslayers} demonstrates that $\ell=128$ is well past the number of layers needed to faithfully sample from a 2-design on each of the symmetry blocks. Thus, in this large layer limit, the scaling with $N_\mathcal{E}$ is independent of the circuit construction, as every block simply represents a random CUE matrix in this limit. Fitting the curves in the lower right panel of Fig.~\ref{fig:scalingvslayers}a ($\ell=128$) and similar curves for $N=4,6,8$ sites and, then, extrapolating to determine the number of samples $N_\mathcal{E}$ needed to reach an error of $\epsilon=0.01$ leads to \Fig{fig:scalingpn}(a).

In the large layer limit, the scaling with Hilbert space dimension $d_s$ observed in this figure applies equally well to symmetry-ignorant schemes if $d_s$ is replaced with the full Hilbert space $d_{\mathcal{H}}$, i.e. the scaling observed is simply a property of sampling from any set of unitaries that forms a unitary $2$-design in the $N_\mathcal{E}\rightarrow\infty$ limit.  We use this fact and the fit in Eq.~(\ref{eq:scaling2design}) to generate the inset of \Fig{fig:scalingpn}(a) showing the relative gain $r_s$ in number of circuit samples for the symmetry-conscious over a symmetry-ignorant scheme.

\subsection{Details for Figure 3(b)}
We now turn to providing additional details for Fig.~\ref{fig:scalingpn}(b), which shows the sample cost scaling for estimating sector-wise $k$-purities for subsystems of size $N_A=N/2$ for a system in the ground state of the particle-number symmetry-preserving Hamiltonian in Eq.~(\ref{eq:HamiltonianModel1}).
Recall, that we consider periodic boundary conditions and systems of size $N=4, 6, 8, 10, 12, 14, 16, 18$ with couplings such that $ma=0.05$ (i.e., a parameter regime where the ground state is entangled). $k$-purities can be extracted via randomized measurement schemes by utilizing the identities in Eq.~(\ref{eq:kpurities}), which hold when the expectation values $\langle P_U(b,s)^k\rangle$ taken with respect to the ensemble of random unitaries $\mathcal{E}$ converge to the value obtained for an ensemble that is an exact unitary $k$-design. 

\begin{figure}
    \centering
    \includegraphics[width=0.95\linewidth]{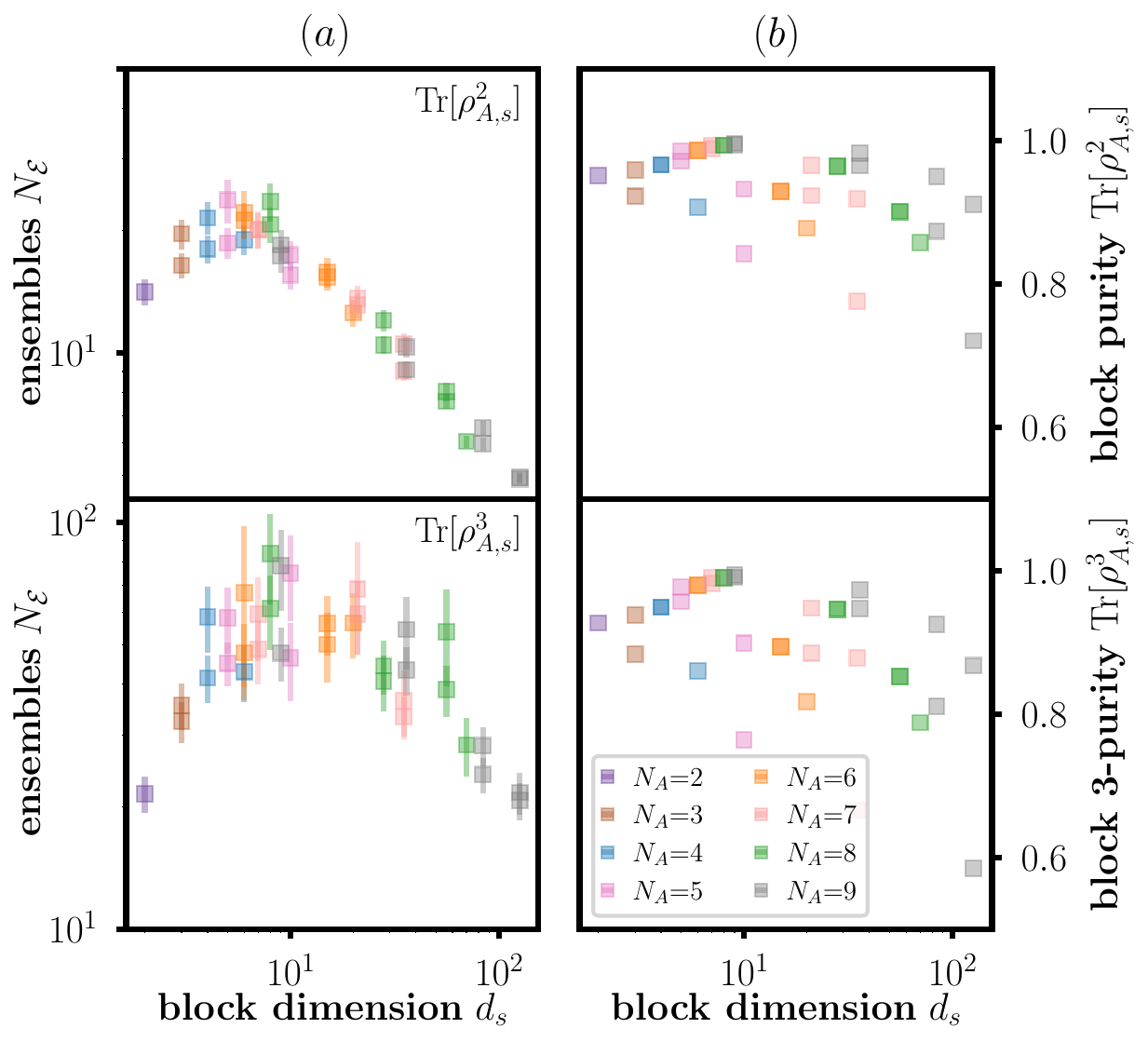}
    \caption{(a) The number of ensembles $N_{\mathcal{E}}$ required, in the infinite shot limit,  to estimate the 2-purity and 3-purity of a single block to 5 percent error. (b) Actual $k$-purities for reduced density matrices on subsystems of size $N_A=N/2$ for each particle number symmetry block for the (normalized) states used in Fig. \ref{fig:scalingpn}(b).}
    \label{fig:True_purities_particle_number}
\end{figure}

A sufficient condition for this convergence is that $\mathcal{E}$ forms an approximate unitary $k$-design; however, note that it is possible that these expectation values converge to a fixed error for a smaller number of samples $N_\mathcal{E}$ than is needed to converge to the same fixed error in being a unitary $k$-design. This is indeed what is observed, as can be seen by comparing the number of samples needed for a convergent estimate of the $k$-purities, shown in Fig.~\ref{fig:True_purities_particle_number}(a), to the number of samples needed to reach an approximate unitary $k$-design, shown in Fig.~\ref{fig:scalingvslayers}(a). This is because an approximate unitary $k$-design will reproduce expectation values of all operators of degree $k$, whereas for this scheme, we must only reproduce the $k$-purities. It is also important to note that the accuracy is inherently dependent on the state under consideration; the $k$-design bound provides a worst case scenario.  

In Fig.~\ref{fig:True_purities_particle_number}(a), we have shown the number of ensembles $N_\mathcal{E}$ required to estimate the 2- and 3-purities to a relative error of $\epsilon = 0.05$ in the infinite shot limit $N_M \to \infty$. The necessary number of ensembles peaks and then begins to decrease as a function of block dimension. This trend cannot continue indefinitely (one must always implement at least one random unitary); consequently, we expect this behavior to saturate for large enough block dimension, as the variance of the infinite shot purity estimator approaches a constant in the large Hilbert space dimension limit~\cite{van2012measuring}.

To create Fig.~\ref{fig:scalingpn}(b) we fix the number of ensembles $N_\mathcal{E} = 1428$, well beyond the the number of ensembles needed to predict the 2-purity and 3-purity to within 5 percent for all cases considered. Since $N_\mathcal{E}$ is constant, this allows us to consider the scaling of the sample cost to purely depend on the number of shots per random unitary from the ensemble (i.e. the number of measurements $N_M$ made in each random basis). Therefore the cost $N_M$ plotted in  Fig.~\ref{fig:scalingpn}(b) is representative of the full sample cost of estimating the $2$-purity of the subsystem states $\rho_A$.

As a sanity check on the $N_\mathcal{E}$ scaling, in Fig.~\ref{fig:True_purities_particle_number}(b) we plot the true $k$-purities for the states used to create Fig.~\ref{fig:scalingpn}(b) and Fig.~\ref{fig:True_purities_particle_number}(a). Note that the states, even for large Hilbert space dimension, have purities of order one. As the cost in $N_\mathcal{E}$ for purity estimation is expected to be largest in the pure state case \cite{van2012measuring}, this shows that the trend in Fig.~\ref{fig:True_purities_particle_number}(a) is not simply because states at large $N$ (large $d_s$) are less pure.

\section{Details for the (1+1)d \texorpdfstring{$\Ztwo$}{Z2} LGT example}\label{app:z21ddetails}
\begin{figure}
    \centering
    \includegraphics[scale=0.44]{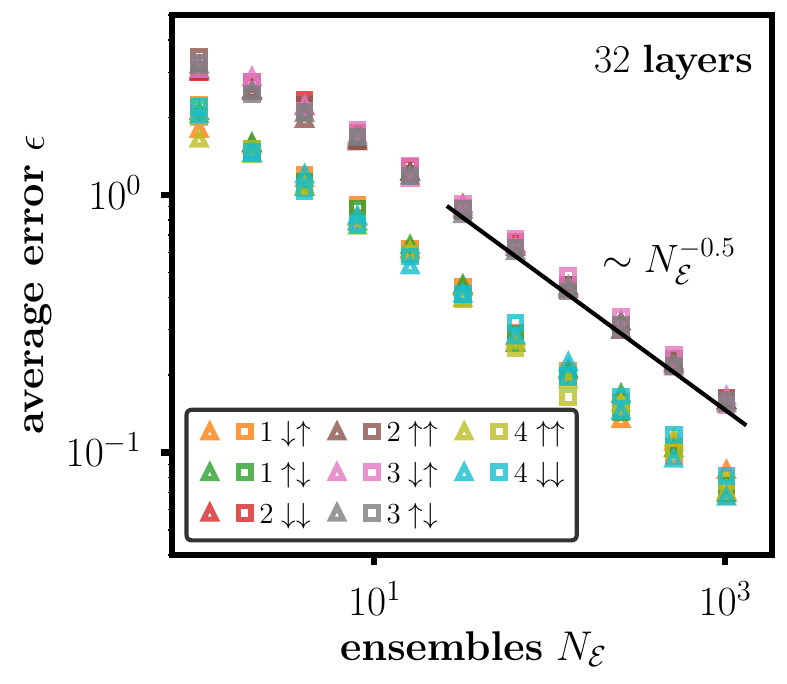}
    \caption{Average error $\epsilon$ in 2-design matrix elements (Eq.~(\ref{eq:def2design})), normalized by $d_s(d_s^2-1)$, versus number of random unitaries $N_\mathcal{E}$  sampled from either the symmetry-respecting random circuits of Section~\ref{ss:1dz2} of depth $\ell=32$ (triangles) or from direct sampling from the CUE within each symmetry sector (squares) for $\mathbb{Z}_{2}^{1+1}$ LGT for a subsystem of size $N_A=5$ ($N=10$). Number of indices sampled is 900.}
    \label{fig:2designtestz21d}
\end{figure}
In the main text, we presented circuits forming symmetry-conscious $k$-designs for $\Ztwo$ LGT in (1+1)d with matter. We demonstrated that these circuits allow to  measure $k$-purities and von Neumann entropies within each symmetry sector, as well as separately extracting the symmetry and distillable entanglement, and the symmetry-resolved Schmidt spectrum using classical shadows. A sufficient, but not necessary condition for such randomized measurement schemes to be successful is that the randomizing circuits form approximate unitary $k$-designs. In this appendix, we explicitly demonstrate that the circuits in question do, in fact, form a sector-wise approximate unitary $2$-design. In particular, we show that they reproduce the correct $2$-design matrix elements (see \Eq{eq:def2design}) for sufficiently deep circuits. 

Representative results are shown in Fig.~\ref{fig:2designtestz21d} for a subsystem of size $N_A=5$ (9 qubits) of a $N=10$ (matter) site system (20 qubits). As described in the main text, Gauss laws at the entanglement boundaries lead to symmetries of  $\rho_A$ in the subsystem. We demonstrate that, within each sector, the random circuits described in Sect.~\ref{ss:1dz2} form a 2-design, by computing the error defined in \Eq{eq:ourerror} with respect to $N_\mathcal{E}$ random circuits with $\ell=32$ layers for all non-trivial symmetry sectors calculated.\footnote{Filling sectors $n_A=0$ and $n_A=N^A$ are trivial as they have unit block size and are not shown.} To compute this error, we average over $|S|=900$ random non-zero matrix elements. We see good agreement between sampling from our random circuits versus sampling directly from the CUE, indicating that our circuits do indeed form approximate unitary 2-designs.

\section{Details of the (2+1)d \texorpdfstring{$\Ztwo$}{Z2} LGT example}\label{app:z22ddetails}
In this appendix, we provide details of the analysis of $\mathbb{Z}_{2}^{2+1}$, discussed in the main text. 

\subsection{Approximate unitary $k$-designs and $k$-purities}
\begin{figure}
    \centering
    \includegraphics[scale=0.44]{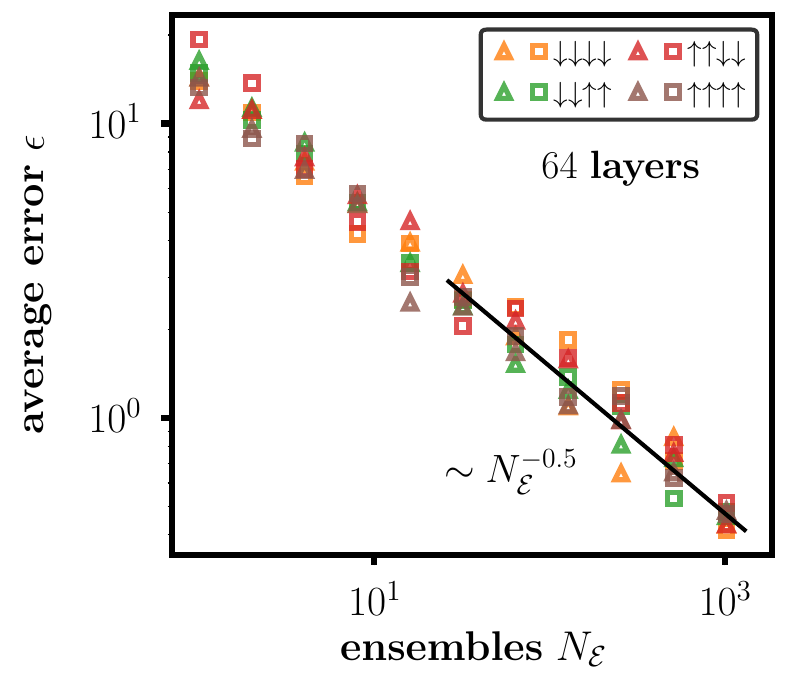}
    \caption{Average error $\epsilon$ in 2-design matrix elements (Eq.~(\ref{eq:def2design})), normalized by $d_s(d_s^2-1)$, versus number of random unitaries $N_\mathcal{E}$  sampled from either symmetry-respecting random circuits of depth $\ell=64$ (triangles) or from direct sampling from the CUE within each symmetry sector (squares) for $\mathbb{Z}_{2}^{2+1}$ LGT for a subsystem of size $3\times 2$ (with fixed boundary conditions in $y$). Number of indices sampled is 900.}
    \label{fig:2designtestz22d}
\end{figure}

The determination of $k$-purities follows exactly that in (1+1) spacetime dimensions. We explicitly show that the (2+1)d circuits in \Fig{fig:2d_setup}(b) explicitly realizes a $2$-design by repeating the analysis of Section \ref{sec:globalsymmetry}. Representative results for a subsystem of size $3\times 2$ are summarized in \Fig{fig:2designtestz22d} showing the error Eq.~(\ref{eq:ourerror}), for every symmetry sector $s$, demonstrating agreement between sampling from our circuits for a sufficient number of layers and sampling directly from the CUE in each sector. Further, we see convergence with increasing samples $N_{\mathcal{E}}$ with the standard $1/\sqrt{N_\mathcal{E}}$ scaling. 

\begin{figure}[t]
  \centering
  \includegraphics[scale=0.44]{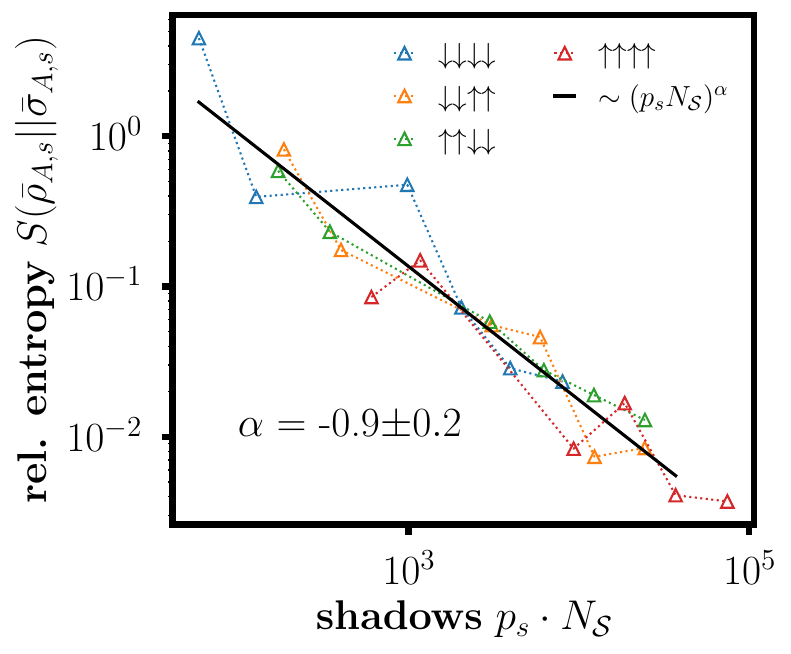}
  \caption{Relative entropy between exact and shadow-reconstructed symmetry-resolved $\rho_{A,s}$ for the $\mathbb{Z}_{2}^{2+1}$ LGT ground state at $\epsilon=0.2$, with $N_x\times N_y=(3+5)\times 2 $ and fixed BC in $y$, $\ell=64$ circuit layers. Data for $N_{\mathcal{S}}=2^{16}$ are shown in \Fig{fig:LGT2dEHT}(b) of the main text.\label{fig:app:KL}}
\end{figure}
\subsection{Classical shadow analysis}
The classical shadow analysis for $\mathbb{Z}_{2}^{2+1}$ ground states follows the previously discussed (1+1)d case. \Fig{fig:app:KL} shows 
the sectorwise relative entropy,
\begin{align}\label{eq:relentr:app}
    S(\bar{\rho}_{A,s}|| \bar{\sigma}_{A,s})\equiv -\text{tr}_s[ \bar{\rho}_{A,s} (\log(\bar{\rho}_{A,s})-\log(\bar{\sigma}_{A,s}) ]
\end{align}
where $\bar{\rho}_{A,s}$ and $\bar{\sigma}_{A,s}$ are the exact and shadow-reconstructed reduced density matrices (projected onto symmetry block $s$) of the $\mathbb{Z}_{2}^{2+1}$ ground state at $\epsilon=0.2$, with $N_x\times N_y=(3+5)\times 2 $, fixed BC in $y$, and $\ell=64$ layers; bars indicate normalization i.e. $\bar{\rho}_{A,s}={\rho}_{A,s}/p_s$ where $p_s = \text{tr}_s[{\rho}_{A,s}]$; $\text{Tr}_s$ denotes the trace over sector $s\in\{ \ua/\da\}^{2^{2N_y}}$. An accuracy of up to $10^{-2}-10^{-3}$ is achieved for the largest samples (where $p_s \cdot N_{\mathcal{S}}\gtrapprox 10^4$); the BW-EHT ansatz at similar cost typically reaches a precision better than $10^{-5}-10^{-6}$ for the same configuration. Convergence with increasing shadow number $N_{\mathcal{S}}$ of the shadow-reconstructed density matrix towards the exact one is evident and shows a power-law behavior consistent with the scaling of the (1+1)d case within error bars.
\begin{figure}[t]
  \centering
  \includegraphics[scale=0.46]{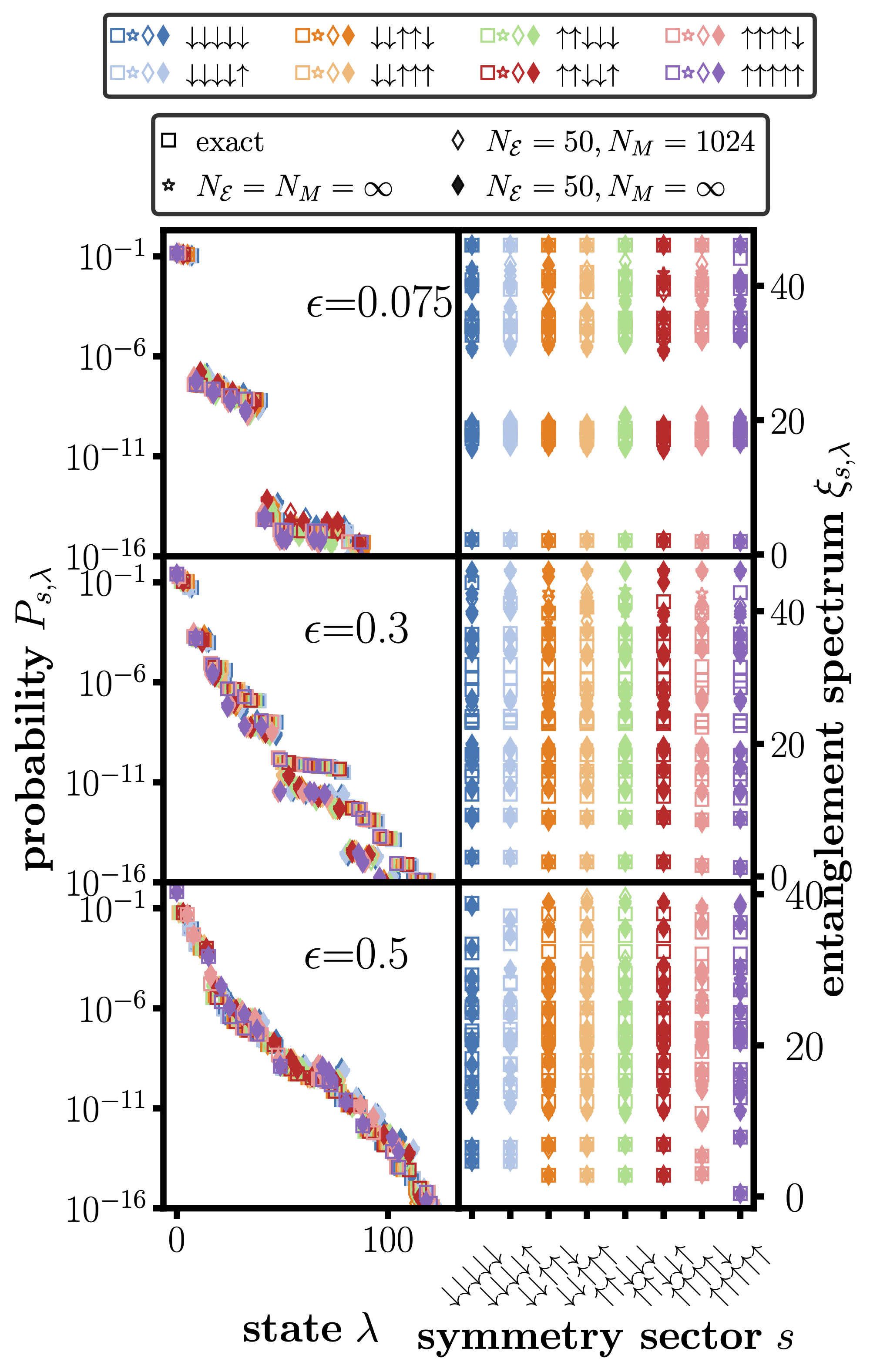}
  \caption{Left column: Symmetry resolved Schmidt spectrum $P_{s,\lambda}$, reconstructed using BW-EHT, for $\epsilon=0.075,0.3,0.5$ and with $N_x\times N_y=(3+3)\times 2$ and periodic boundary conditions in $y$ (and $x$),  $\ell=64$, $N_{\mathcal{E}}=50$, $N_{M}=1024$. Right column: Symmetry resolved Entanglement spectrum.\label{fig:app:BW2d}}
\end{figure}

\subsection{Entanglement Hamiltonian tomography analysis}
Finally, our Bisognano-Wichmann theorem based entanglement Hamiltonian tomography (BW-EHT) protocol follows Ref.~\cite{kokail2021entanglement}, except that we perform the optimization in every symmetry-sector $s$ separately. The approach is based on an alternative representation of a reduced density matrix with Schmidt representation $\rho_A = \sum_{\lambda} P_{\lambda} | \lambda \rangle \langle \lambda|$. In particular, one defines an Entanglement Hamiltonian as
\begin{align}
H_A = -\log[\rho_A].
\end{align}
Eigenvalues are given by $P_\lambda=\exp(-\xi_\lambda)$, where $\xi_\lambda$ are the eigenvalues of $H_A$. 

Because $\rho_A$ can be split into its corresponding symmetry sectors as $\rho_A = \bigoplus_s \rho_{A,s}$ it follows that also $H_A = \bigoplus_s H_{A,s}$, where $s$ labels quantum numbers of the spectrum $\xi_{\lambda,s}$ of $H_{A,s}$. Inserting Eq.~(\ref{eq:BW}), for the Entanglement Hamiltonian we obtain the following BW-EHT ansatz for the state within symmetry sector labeled by quantum number $s$:
\begin{align}
\bar{\rho}_{A,s}[\{ \beta_i\}]\equiv \frac{\exp\{ - H_{A,s}[\beta_i]\}}{\text{Tr}_s[\exp\{ - H_{A,s}[\beta_i]\} ]}
\end{align}
where $H_{A,s}$ is is a deformation of the physical Hamiltonian, i.e.  local couplings $\beta_i$ replace the physical couplings. The ansatz is such that the state is normalized within each symmetry sector, i.e., $\text{Tr}_s[\bar{\rho}_s] = \text{Tr}_s[{\rho}] / p_s=1  $, where $p_s$ is the probability of being in sector $s$. We find the optimal couplings $\{\beta_i\}$ by minimizing the following functional
\begin{align}\label{eq:chi2}
 \sum_{b} \Big\langle \Big(  P_U(b,s) - \text{Tr}_s \big[  \bar{\rho}_{A,s}U_s | b,s \rangle \langle b,s | U^\dagger_s  \big] \Big)^2\Big \rangle_{ \mathcal{E}},
\end{align}
separately for each $s$. Here, $ \mathcal{E} = \{ U_s\}$ is the ensemble of random circuits restricted to the block labeled by $s$, and $P_U(b,s)$ is the probability of measuring outcome bit string $b$, normalized such that $\sum_b P_U(b,s)=1$ for all $s$. $P_U(b,s)$ is determined   by classically simulating an (ideal) circuit for a given number of shots. In practice, the optimization is performed using python's simplicial homology global optimization (scipy.optimize.shgo)~\cite{scipyref} with the following parameters
\begin{verbatim}
scipy.optimze.shgo(chi_squared, bounds,n=32,
sampling_method='sobol', options=opt_dict)
\end{verbatim}
with sampling method `sobol', and $n=32$ sampling points in the construction of the simplicial complex, and very large bounds i.e. typically $ \beta_i \in [-30.0,30.0]$; all other options are set to their default values.

An example of the results of this analysis is shown in \Fig{fig:app:BW2d}, where we show the symmetry-resolved Schmidt  $P_{s,\lambda}$ and entanglement spectrum $\xi_{\lambda,s}$ of the $\mathbb{Z}_{2}^{2+1}$ ground state at $\epsilon=0.075,0.3,0.5$, for $N_x\times N_y=(3+3)\times 2$, periodic boundary conditions in $y$ and  $\ell=64$. To estimate the error from applying a finite number of random circuits and estimating the effect of shot noise on obtaining $P_U(b,s)$, we additionally perform the following analysis:
We compute the exact state using exact diagonalization. We then numerically minimize, within each symmetry sector $s$, the relative entropy between the exact density operator $\bar{\sigma}_s$, normalized so that $\mathrm{Tr}_s[\bar\sigma]=1$, and the BW ansatz $\bar\rho_s$,
\begin{align}\label{eq:BW2}
  S(\bar{\sigma}_s || \bar{\rho}_s) &\equiv {\text{Tr}}_s [\bar{\sigma}_s( \log(\bar{\sigma}_s) - 
\log(\bar{\rho}_s))] \nonumber\\
&= -S(\bar{\sigma}_s) + \Gamma(\bar{\rho}_s || \bar{\sigma}_s) \ge 0\,,
\end{align}
$S(\bar{\sigma}_s)$ is the exact von Neumann entropy and
\begin{align}\label{eq:gamma6}
\Gamma(\bar{\rho}_s|| \bar{\sigma}_s) \equiv \log(\text{Tr}_s [\bar{\rho}_{A,s}])+\sum_i \beta_i \text{Tr}_s[H_i \bar{\sigma}_s]),.
\end{align}
 can be easily computed. Minimizing \Eq{eq:BW2} with the exact same numerical optimization as used for \Eq{eq:chi2} provides the BW-EHT result in the infinite measurement bases and infinite shot limit. The deviation of our circuit simulation  from this result provides the  error that we show in Fig.~\ref{fig:LGT2dEHT}(c).

\bibliography{references.bib}

\end{document}